 \documentclass[preprint,11pt]{elsarticle}

\usepackage{amssymb}
\usepackage{amsmath}
\usepackage[utf8]{inputenc}
\usepackage{graphicx}
\usepackage{float}
\usepackage{epstopdf}
\usepackage{csquotes}
\usepackage{amsmath}
\usepackage{subcaption}
\usepackage{caption}
\usepackage[T1]{fontenc}
\usepackage[linesnumbered,ruled]{algorithm2e}
\usepackage{hyperref}
\graphicspath{{./Fig/}}
\journal{}

\begin{document}

\begin{frontmatter}



\title{On complexity of post-processing in analyzing GATE-driven X-ray spectrum}


\author[lbl1]{Neda Gholami  \corref{cor1}} \ead{gholami@iranprc.org}
\author[lbl1]{Mohammad Mahdi Dehshibi \corref{cor2}} \ead{dehshibi@iranprc.org, mohammad.dehshibi@yahoo.com}
\author[lbl2]{Mahmood Fazlali}
\author[lbl2b,lbl2b2,lbl3b]{Antonio Rueda-Toicen}
\author[lbl2b,lbl3a,lbl3b]{Hector Zenil}
\author[lbl4]{Andrew Adamatzky}

\address[lbl1]{Pattern Research Center, Iran}
\address[lbl2]{Department of Computer Science, Shahid Beheshti University, G.C., Tehran, Iran}
\address[lbl2b]{Algorithmic Nature Group, LABORES for the Natural and Digital Sciences, Paris, France}
\address[lbl2b2]{Instituto Nacional de Bioingenier\'{i}a, UCV, Caracas, Venezuela}
\address[lbl3a]{Oxford Immune Algorithmics, Oxford University Innovation, Oxford, U.K.}
\address[lbl3b]{Algorithmic Dynamics Lab, Unit of Computational Medicine, SciLifeLab, Centre for Molecular Medicine, Department of Medicine Solna, Karolinska Institute, Stockholm, Sweden.}
\address[lbl4]{Department of Computer Science, University of the West of England, Bristol, U.K.}

\cortext[cor1]{Corresponding author}
\cortext[cor2]{Principal Corresponding author}

\begin{abstract}
\textit{Purpose:} Computed Tomography (CT) imaging is one of the most influential diagnostic methods. In clinical reconstruction, an effective energy is used instead of total X-ray spectrum. This approximation causes an accuracy decline. To increase the contrast, single source or dual source dual energy CT can be used to reach optimal values of tissue differentiation. However, these infrastructures are still at the laboratory level, and their safeties for patients are still yet to mature. Therefore, computer modelling of DECT could be used.\\
\textit{Methods:} We propose a novel post-processing approach for converting a total X-ray spectrum into irregular intervals of quantized energy. We simulate a phantom in GATE/GEANT4 and irradiate it based on CT configuration. Inverse Radon transform is applied to the acquired sinogram to construct the Pixel-based Attenuation Matrix (PAM). To construct images represented by each interval, water attenuation coefficient of the interval is extracted from NIST and used in the Hounsfield unit (HU) scale in conjunction with PAM.  The CT image is modified by using of an associated normalized photon flux and calculated HU corresponding to the interval. \\
\textit{Results:} We demonstrate the proposed method efficiency via complexity analysis,  using absolute and relative complexities, entropy measures, Kolmogorov complexity, morphological richness, and quantitative segmentation criteria associated with standard fuzzy C-means.\\
\textit{Conclusions:} The irregularity of the modified CT images decreases over the simulated ones.
\end{abstract}


\end{frontmatter}


\section{Introduction}

Clinical imaging techniques, e.g., radiology, mammography, X-ray computed tomography (CT), magnetic resonance imaging (MRI), single-photon emission computed tomography (SPECT), are key component of medical diagnostics. CT is the most widely used technique in which attenuating properties of different tissues such as fat, bones, and muscles, are used to visualize of each voxel and the associated Hounsfield Unit (HU). These tissues have known X-ray attenuation coefficients which are used as the basis of data acquisition \cite{ginat2014advances}, \cite{ginat2015computed}. The CT image is affected by a scanner type, projection systems, and reconstruction algorithms~\cite{cierniak2011x}. Due to the costs of physical development of the scanners and unnecessary rapid patient exposure, a substantial domain of work is implemented in computational models ~\cite{duan2017computed}. Mah et al.~\cite{mah2010deriving} investigated a relationship between grey levels of images and Hounsfield units (HU) in cone beam CT (CBCT) scanners. It was reported that there exist a linear relationship between the grey levels and the attenuation coefficient of each of the materials at \enquote{effective} energy. Linearity was proved by calculating the linear regression of attenuation coefficients for the reference materials. A negligible difference could be found between actual Hounsfield units of each phantom material at the selected effective energy and those calculated from grey levels.

The projection sub-systems of CT scanners has experienced changes in 3 aspects including parallel, fan, and cone beam systems \cite{sidky2008image}-\cite{zhihua2008direct}. Sidky and Pan \cite{sidky2008image} proposed a theoretical framework, namely total variation (TV), to show how accurate circular cone-beam CT image reconstruction can be done from reduced data sampling. They argued that TV algorithm can resolve low-contrast structures in the presence of high-contrast objects. Zhihua and Guang-Hong \cite{zhihua2008direct} considered image reconstruction in fan-beam based CT. This method is subject to observing full circle scan in data acquisition mode. Experimental setup demonstrated that when the image object is relatively large, the fan angle must increase to cover the entire image object, and a parallel-beam approximation cannot be directly applied to reconstruct images.

Image reconstruction in CT is an inverse problem which can be categorized into two categories, namely analytical reconstruction and iterative reconstruction \cite{herman2009fundamentals}. In the former category, there can be found many algorithms among which filtered back-projection (FBP) \cite{mouton2013experimental} is the most acceptable one. FBP, which is derived by using the Fourier Slice Theorem \cite{ng2005fourier}, uses a 1D filter prior to back-projecting data into the image space. This method is computationally efficient and has numerical stability. In the latter category, however, statistical, likelihood-based iterative expectation-maximization algorithms \cite{lange1984reconstruction}, \cite{vardi1985statistical} are preferred methods. These algorithms estimate the probability distribution of annihilation events that led to the measured data. The advantages of the iterative approach are insensitive to noise, the capability of reconstructing an optimal image in the case of incomplete data, and resistance to the streak artefacts common with FBP \cite{wang1996iterative}. This category of methods, which is alternatively known as algebraic methods, has been applied in emission tomography modalities, e.g., SPECT and PET where the attenuation along ray paths is significant, and noise statistics are relatively poor \cite{boas2011evaluation}.

Majority of computational methods aim at using the attenuation coefficients in the effective energy of the total X-ray spectrum instead of the real one to ease the reconstruction CT image. Although some methods could cover a broader range of energies, they need multi-irradiation in clinical levels with the associated risks. Besides, using a specific energy, as an alternative to the total energy, leads to decreasing of contrast level and, in turn, the accuracy. Therefore, establishing a trade-off among accuracy, decreasing irradiation defects, and computational cost is especially considered in this research. As far as we know, the idea of spanning effective energy was first proposed in \cite{gholamiCT2015}. However, there are two critical technical issues with the implementation of the idea. First, the inverse HU was applied to the reconstructed image obtained from back-projecting the HU in the energy level of 70 keV to form the attenuation map. Second, energy quantization was done without considering the statistical distribution of the source photon flux. The first issue causes the rest of analysis were done on a back-projected data where the effect of applying HU were neutralized with the inverse HU. Moreover, the raw quantization led to increasing nonsense data and reducing the accuracy of calculations. In this study, we also contribute towards the role of post-processing in reconstructing of the total X-ray spectrum by covering more energy range in the computational level as well as resolving the mentioned technical issues.

Simulating a phantom, which is irradiated by an X-ray source, is the primary prerequisite to validate our hypothesis in this study. Constructed phantom consists of three rectangular cubes made of a skull, rib bone, and lung tissues surrounded by a water cylinder in GATE/GEANT4 environment. Two main reasons for considering these tissues are (1) attenuation coefficients of water are close to the lung, and the same condition exists for the rib bone and skull, and (2) discrimination among tissues is large enough which makes the experiments fair. The radiation source is set to the range of 0-140 keV fan-beam X-ray in a way that could cover double-wedge.

The back-projection method is applied to the irradiated phantom to reconstruct images, so-called pixel-based attenuation matrix (PAM), in which the inverse radon transform \cite{cabral1994accelerated} is utilized. Since the attenuation coefficient of each tissue is different, a specific value should be, then, calculated in each energy level. To make the image representation as simple as possible, the total X-ray spectrum is replaced by the effective energy and the corresponding water attenuation coefficient where the calculated HU could bring a higher intensity representation (refer to Eq. \ref{eq:1}).

\begin{equation}
HU = \frac{\mu - \mu_{w}}{\mu_{w}} \times 1000
\label{eq:1}
\end{equation}

where $\mu$ is the attenuation coefficient, $\mu_{w}$ is the water attenuation coefficient, and HU is Hounsfield unit scale. Effective energy is usually set to 70 keV for the X-ray spectrum with the energy variation of 0-140 keV. Producing of mono-energetic images in clinical data acquisition are subject to solving non-linear equations which is not computationally feasible. In this study, we contribute towards the quantization of the X-ray spectrum by mapping the acquired data to 13 irregular intervals. In the line of calculations, those energy values which were lower than 10 keV were considered as outliers and overlooked accordingly. To calculate the effective energy of each interval by using Eq. \ref{eq:1}, first, the statistical average energy of the interval in conjunction with the known water attenuation coefficient (refer to Table \ref{tbl:1}) are used. Then, the value of HU is weighted by using the calculated PAM and associated normalized photon flux to that interval. It was observed in the course of experiments that the proposed method would increase the contrast of target tissue in a specific energy interval where it is not visible in another energy. Meanwhile, it can reduce the complexity of CT images for the further analysis in the segmentation task. It is because the different attenuation coefficients of different tissues which force the radiologist to irradiate the patient repeatedly if an exact diagnosis is desirable.

\begin{table}[H]
	\centering
	\caption{Values of the mass attenuation coefficient, $\mu/\rho$ as a function of photon energy for water \cite{international1989tissue}.}
	\label{tbl:1}
	\begin{tabular}{ll}
		\hline
		Energy (keV) & $\mu/\rho$ ($cm^{2}/g$) \\ \hline
		1.00000 E+01  & 5.329 E+00               \\
		1.50000 E+01  & 1.673 E+00               \\
		2.00000 E+01  & 8.096 E-01               \\
		3.00000 E+01  & 3.756 E-01               \\
		4.00000 E+01  & 2.683 E-01               \\
		5.00000 E+01  & 2.269 E-01               \\
		6.00000 E+01  & 2.059 E-01               \\
		8.00000 E+01  & 1.837 E-01               \\
		1.00000 E+02  & 1.707 E-01               \\
		1.50000 E+02  & 1.505 E-01               \\ \hline
	\end{tabular}
\end{table}

In brief, the proposed method consists of several main steps: 
(1) back-projecting acquired data to form pixel-based attenuation matrix (PAM), 
(2) finding the statistical average of each interval to use as the effective energies, 
(3) calculating HU scale of each interval, 
(4) computing the associated photon fluxes based on X-ray spectrum, 
(5) modifying HU scales by weighting them with the computed fluxes. The proposed post-processing method is tested using  visual evaluation and complexity analysis. Complexity criteria applied include various entropy measures and Kolmogorov complexity. Absolute and relative complexities, morphological richness as well as quantitative segmentation criteria associated with standard fuzzy C-means are also reported demonstrating that irregularity of the modified CT images decreases over the simulated ones.

The rest of this paper is organized as follows: Section 2 is dedicated to the proposed method. The experimental setup is described in Section 3. Finally, this paper concludes in Section 4.

\section{Methodology}

Multiple irradiating patients is still an issue which we take it in this study by shifting physical procedures into post-processing. Our contributions is founded based on GATE/GEANT4 simulations, which its configuration as well as the details of post-processing approach are described here.

\subsection{Simulating X-ray spectrum in GATE/GEANT4}

A phantom is created and then radiated in GATE. We define a coordinate system as a cube of air with $50$ cm: 

\begin{equation}
\begin{bmatrix}
-25 & +25 & +25 & -25 & -25 & +25 & +25 & -25 \\
+25 & +25 & +25 & +25 & -25 & -25 & -25 & -25 \\
+25 & +25 & -25 & -25 & +25 & +25 & -25 & -25
\end{bmatrix}
\end{equation}

\begin{figure}[H]
	\centering
	\begin{subfigure}[b]{0.45\textwidth}
		\includegraphics[width=\textwidth]{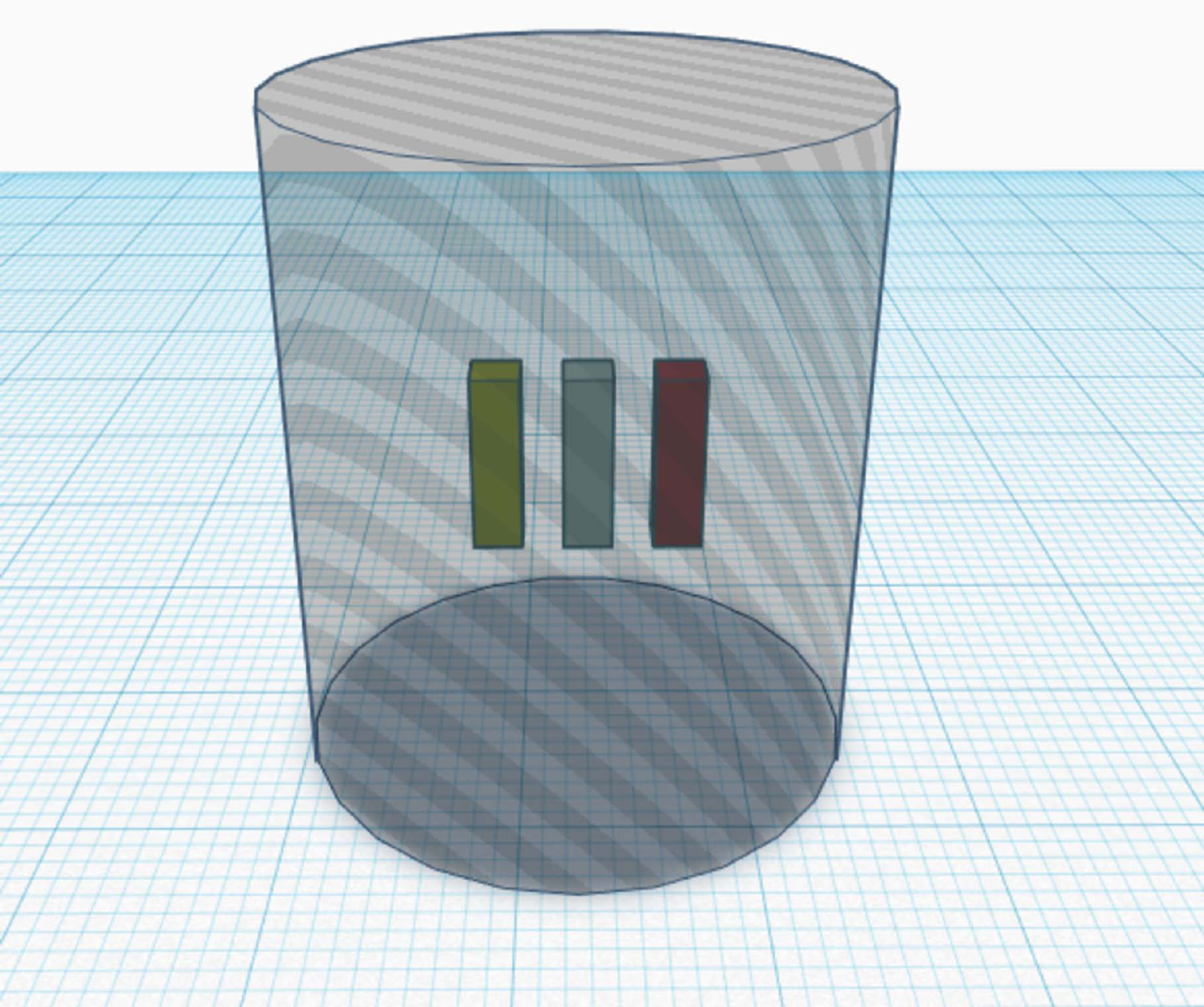}
		\caption{}
	\end{subfigure}
	\hfill
	\begin{subfigure}[b]{0.50\textwidth}
		\includegraphics[width=\textwidth]{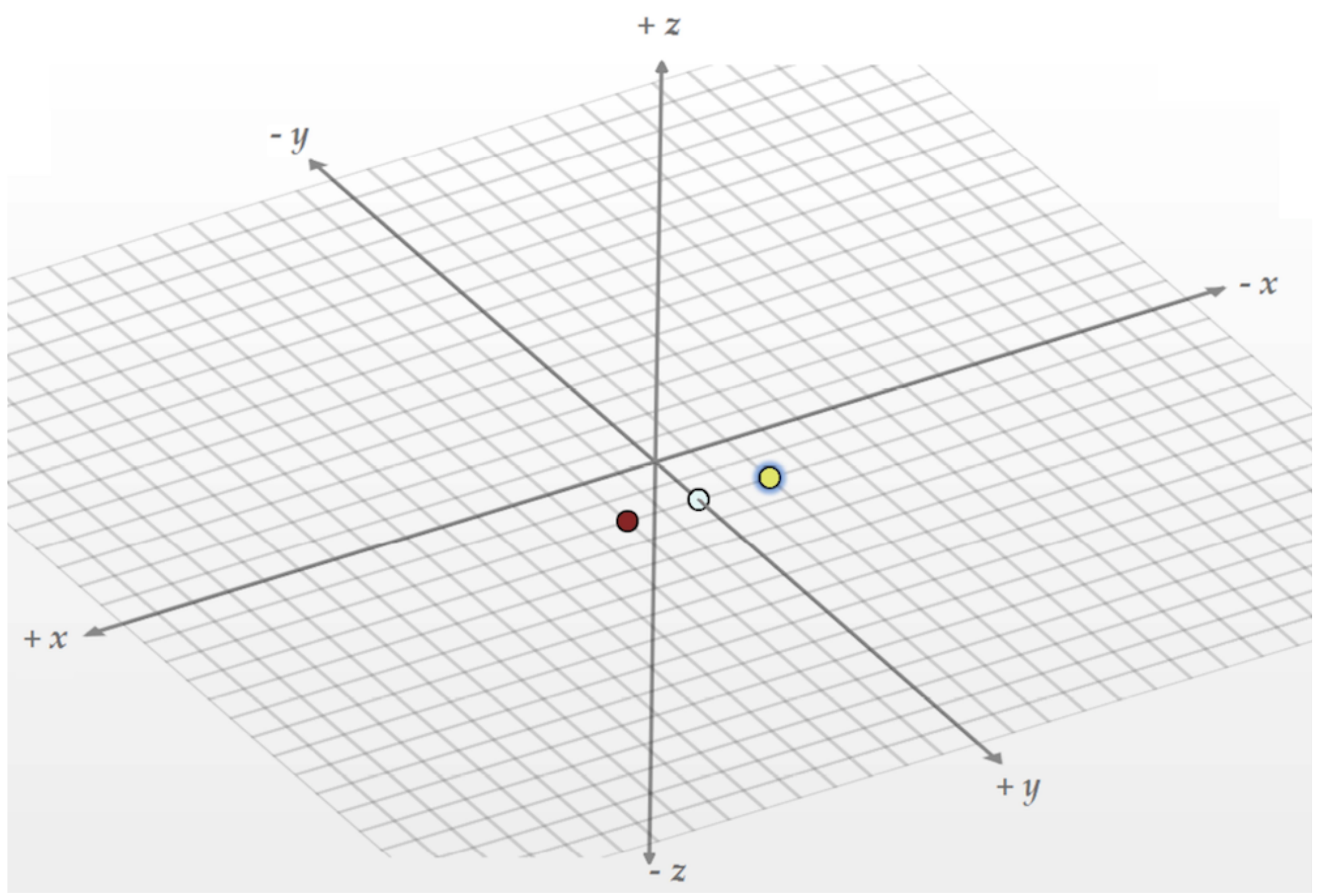}
		\caption{}
	\end{subfigure}
	\caption{(a)~Constructed phantom in GATE. (b)~Coordinates of tissues are (2, 2, 0), (0, 2, 0), and (-2, 2, 0) mm, respectively. Red, white, yellow colors show rib bone, lung, and skull, respectively.}\label{fig:1}
\end{figure}

CT scanner is made of $30 \times 16$ detector arrays which are position  in $ (0, 0, 150.5)$ mm relatively to the defined subspace. Each cell is also a cube with the size of $ 0.5 \times 0.5 \times 1$ mm and made of LSO, i.e., Lutetium, Silicon, and Oxygen. The phantom is a cylinder with the radius of 5 mm and the height of 6 mm includes lung, rib bone, and skull tissues surrounded by water. The size of each tissue is $1 \times 1 \times 2$ mm. The choice of tissue composition and parameters is based on the facts that  (1) attenuation coefficients of water are close to the lung, and the same condition exists for the rib bone and skull, and (2) discrimination among tissues is large enough to make the experiments realistic. Density of tissues are $ 0.26, 1.92$, and $1.61 g/cm^{3}$, respectively. Structure of phantom and positions of tissues are shown in Fig.~ \ref{fig:1}.


The source is a rectangle fan-beam with the size of $0.5 \times 0.5$ mm and placed in the $(0, 0, -150)$ mm. Its activity is set to 100 MBq. In defining this source, the following constraints are taken into account.

\begin{enumerate}
	\item The fan divergence angle $(\theta)$ is set to 6.8 degrees which can cover a surface area of 77.70 mm in the cross point. The coverage area is calculated by $2 \tan(\tfrac{\theta}{2}) \times |Dist_{s}- \frac{H_{p}}{2}|$, where $Dist_{s}$ is the distance between source and phantom and $H_{p}$ is the height of phantom.
	\item The most common activity levels used in laboratories are the millicurie $(mCi)$ and microcurie $(\mu Ci)$, which is equal to $3.7 \times 10^{10} \space Bq$. Here, $Dist_{s}$ and $\theta$ are defined in a way that our simulation can conform to the safety condition of the real imaging setup.
	\item The energy level is in the range of 10-140 keV.
	\item The phantom is defined in a way that it has no activity, i.e., a cold material.
	\item A 360 degrees rotation is desirable. Therefore, in this study, the phantom is rotated over the $z$ axis by 1 degree per second.
\end{enumerate}

\subsection{The proposed post-processing method}

\begin{figure}[H]
	\centering
	\includegraphics[width=0.5\textwidth]{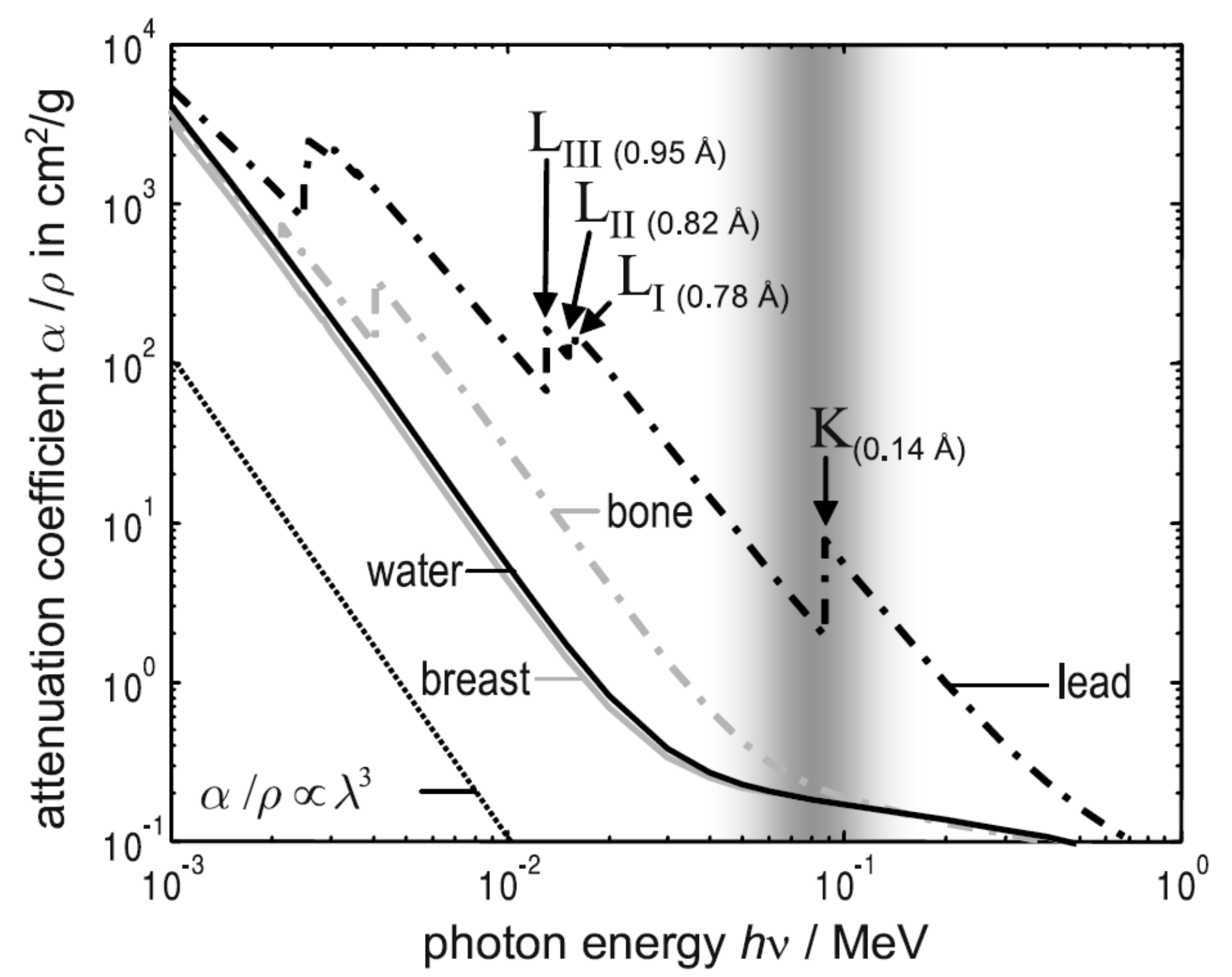}
	\caption{Mass attenuation coefficient ($\alpha/\rho$ measured in units of $cm^{2}/g$) for lead and water as well as for the bio-tissues bone and soft tissue given versus the incident radiation energy. \cite{buzug2008computed}}\label{fig:2}
\end{figure}

Attenuation coefficients of tissues are different and finding the most suitable effective energy for each tissue could, in turn, increase the level of contrast which finally helps physicians to do a better diagnosis. This idea was previously investigated in \cite{buzug2008computed} and Fig.~ \ref{fig:2} can demonstrate it clearly.

The proposed approach is comprised of steps: 
(1) back-projecting acquired data to form pixel-based attenuation matrix (PAM); 
(2) finding the statistical average of each interval to use as the effective energies; 
(3) calculating HU scale of each interval; 
(4) computing the associated photon fluxes based on X-ray spectrum; 
(5) modifying HU scales by weighting them with the computed fluxes. 
Details of this post-processing approach are shown in Algorithm \ref{alg:1}

\begin{algorithm}
	\SetKwInOut{Input}{Input}
	\SetKwInOut{Output}{Output}
	
	\Input{\begin{itemize}
			\item[] PAX $\gets$ projected attenuation X-ray.
			\item[] W $\gets$ water attenuation coefficient.
			\item[] F $\gets$ photon flux value.
	\end{itemize}}
	\Output{wHU $\gets$ weighted HU, known as enhanced CT image.}

	PAM = iradon(PAX) \\
	Form intervals as: \\
	\ \ \ \ \ \ \ \  X $\gets$ \{(12-17), (18-27), (28-37), (38-47), (48-57), (58-67), (60-72), (68-80), (78-87), (81-95), (88-100), (98-105), (130-150)\} \\
	Take Kolmogorov-Smirnov test to find the best distribution that fits X: \\
	\ \ \ \ \ \ \ \ $F_{n}(x)=\tfrac{1}{n}\sum_{i=1}^n I_{[\infty, x]}(X_{i})$, \\
	\ \ \ \ \ \ \ \ $D_{n}=\sup_{x}|F_{n}(x)-F(x)|$ \\
	where $F(x)$ is the hypothesis distribution, $F_{n}(x)$ is the cumulative distribution function, and $I_{[\infty, x]}(X_{i})$ is the indicator function. \\
	Calculate the \enquote{effective energy}:\\
	$\mu_{w} \gets \{\mu_{w_{i}} \mid \mu_{w_{i}} = \mathrm{E}[x], \quad x \in \mathrm{X_{i}}, 1 \leq i \leq 13\}$\\
	$mF=\sum_{i=1}^{13} F_{i}$ \\
	$i = 1$ \\
	\While{$i \leq 13$}{
		$HU_{i} = \frac{PAM-\mu_{w_{i}}}{\mu_{w_{i}}} \times 1000$, \newline
		$q_{i} = \frac{F_{i}}{mF}$, \newline
		$wHU = q_{i} \times HU_{i}$
	}
	\caption{Proposed post-processing algorithm.} \label{alg:1}
\end{algorithm}

\begin{figure}[H]
	\centering
	\begin{subfigure}[b]{0.49\textwidth}
		\includegraphics[width=\textwidth]{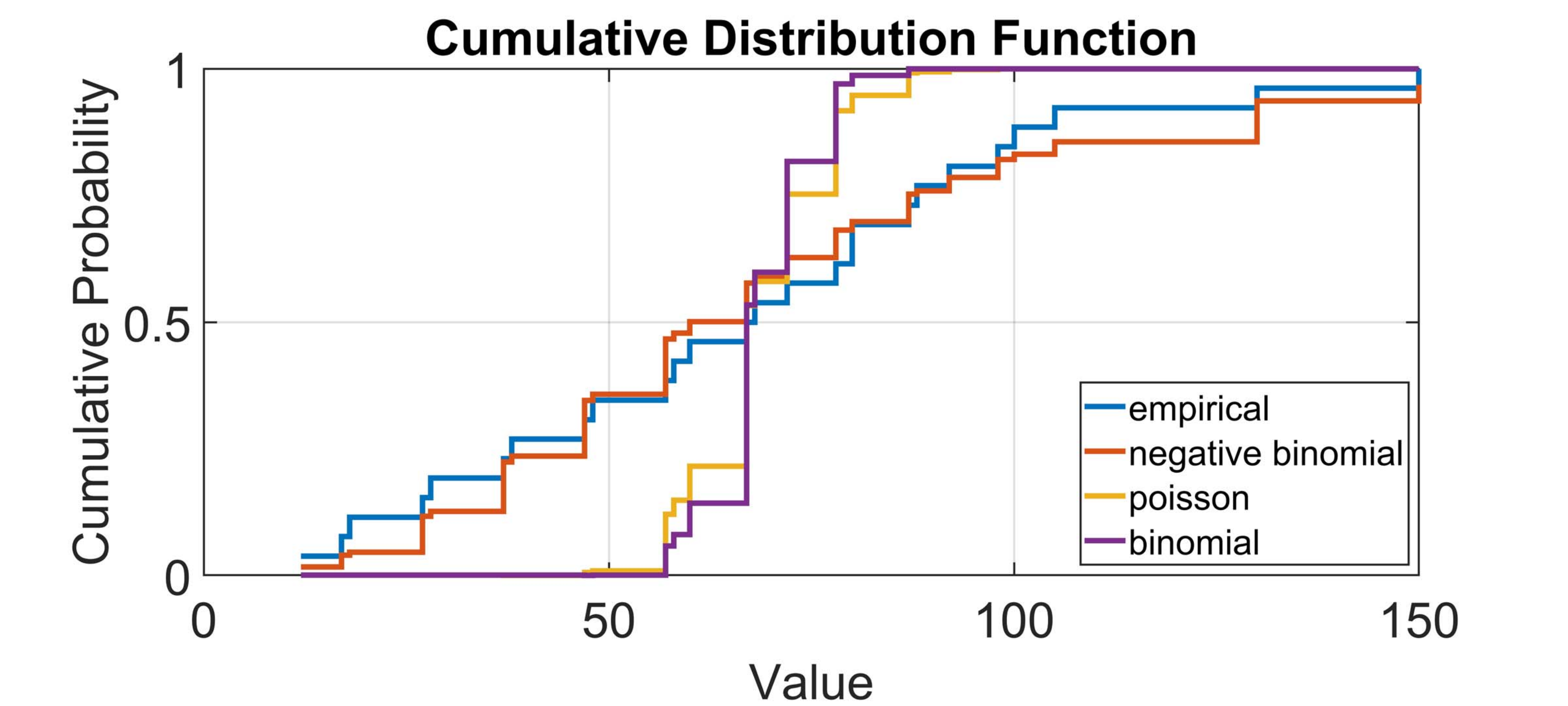}
		\caption{}
	\end{subfigure}
	\hfill
	\begin{subfigure}[b]{0.49\textwidth}
		\includegraphics[width=\textwidth]{Fig_EPS/Fig3a}
		\caption{}
	\end{subfigure}
	\caption{(a) Cumulative density function, and (b) probability density function of endpoints associated to the defined intervals}\label{fig:3}
\end{figure}

The X-ray spectrum of the acquired data is mapped into 13 irregular intervals. Although this sort of quantization roots in the known water attenuation coefficients (WAC) \cite{international1989tissue}, there is no measured WAC for the energy levels of 70 and 95 keV. Therefore, two overlapped intervals are considered in this study for energy levels of 70 and 95 keV to make this approximation as accurate as possible. It should be noted that values with the energy level lower than 10 keV are considered as outliers and overlooked accordingly. To compute the effective energy of each interval, the maximum likelihood estimation (MLE) method is applied to the set of interval's endpoints, $ep$, in order to find the best distribution fits our data. The Kolmogorov-Smirnov test is used \cite{stephens1974edf} to compare the histogram of data to the probability density function. The theoretical cumulative density function (CDF) and probability density function (PDF) are compared to the empirical ones. Results of this test are illustrated in Fig.~ \ref{fig:3}

As shown in Fig.~ \ref{fig:2}, negative binomial distribution \cite{degroot2012probability} can model the  data presented. A number of failures $r$ and success probability, in each experiment, $p$, are calculated to measure the mean of each interval. This mean is considered to be the effective energy of that interval (see Eq. \ref{eq:2a}). HU scale of each effective energy is calculated using Eq. \ref{eq:1}. A modified CT image is constructed by weighting the HU through the normalized photon flux of the interval.

\begin{gather} \label{eq:2a}
mean = \frac{p \cdot r}{1-p}, \\
eeValue = \lfloor (ep_{2}-ep_{1}) \times mean \times 10 \rfloor + ep_{1}.
\end{gather}

\section{Experimental Setup}

Simulating a phantom irradiating by an X-ray source is the necessary component to validate our hypothesis. The back-projection method is applied to the irradiated phantom to reconstruct images, so-called pixel-based attenuation matrix (PAM), in which the inverse radon transform \cite{cabral1994accelerated} is utilized. The proposed post-processing method is tested in the line of segmentation task as well as complexity measures. The approach is validated using a trade-off between accuracy, decreasing irradiation defects, and computational cost. We also analyse a role of post-processing in reconstruction of the total X-ray spectrum by considering more energy intervals at the same computational level.  

\subsection{Simulated CT Data}

\begin{figure}[H]
	\centering
	\begin{subfigure}[b]{0.17\textwidth}
		\includegraphics[width=\textwidth]{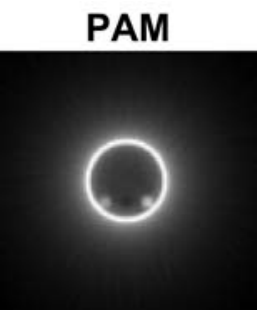}
		\caption{}
	\end{subfigure}
	\hfill
	\begin{subfigure}[b]{0.17\textwidth}
		\includegraphics[width=\textwidth]{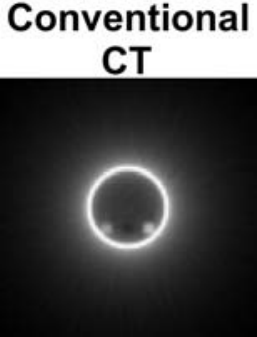}
		\caption{}
	\end{subfigure}
	\hfill
	\begin{subfigure}[b]{0.17\textwidth}
		\includegraphics[width=\textwidth]{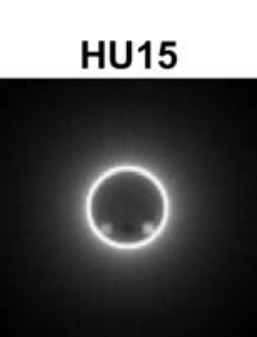}
		\caption{}
	\end{subfigure}
	\hfill
	\begin{subfigure}[b]{0.17\textwidth}
		\includegraphics[width=\textwidth]{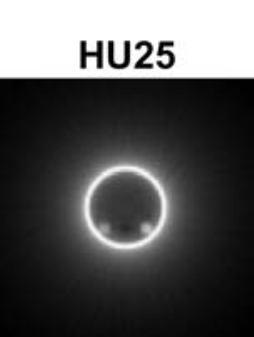}
		\caption{}
	\end{subfigure}
	\hfill
	\begin{subfigure}[b]{0.17\textwidth}
		\includegraphics[width=\textwidth]{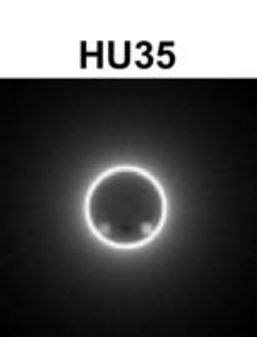}
		\caption{}
	\end{subfigure}
	\hfill
	\begin{subfigure}[b]{0.17\textwidth}
		\includegraphics[width=\textwidth]{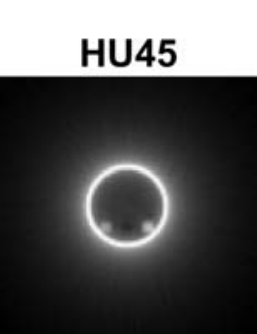}
		\caption{}
	\end{subfigure}
	\hfill
	\begin{subfigure}[b]{0.17\textwidth}
		\includegraphics[width=\textwidth]{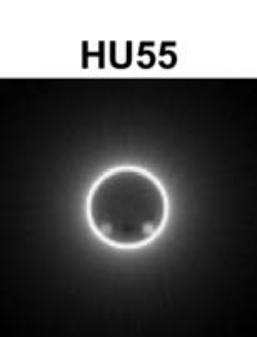}
		\caption{}
	\end{subfigure}
	\hfill
	\begin{subfigure}[b]{0.17\textwidth}
		\includegraphics[width=\textwidth]{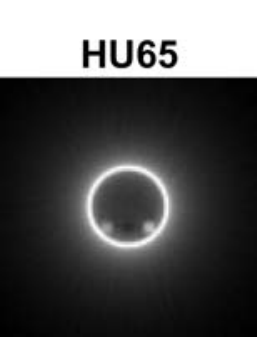}
		\caption{}
	\end{subfigure}
	\hfill
	\begin{subfigure}[b]{0.17\textwidth}
		\includegraphics[width=\textwidth]{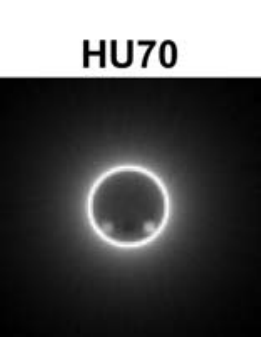}
		\caption{}
	\end{subfigure}
	\hfill
	\begin{subfigure}[b]{0.17\textwidth}
		\includegraphics[width=\textwidth]{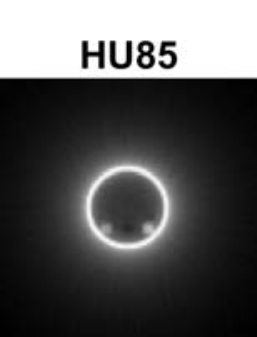}
		\caption{}
	\end{subfigure}
	\hfill
	\begin{subfigure}[b]{0.17\textwidth}
		\includegraphics[width=\textwidth]{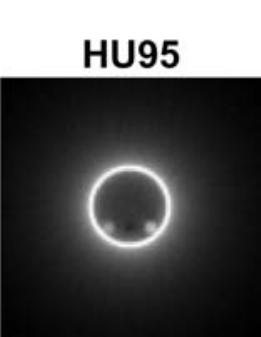}
		\caption{}
	\end{subfigure}
	\hfill
	\begin{subfigure}[b]{0.17\textwidth}
		\includegraphics[width=\textwidth]{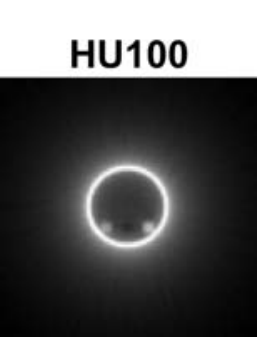}
		\caption{}
	\end{subfigure}
	\hfill
	\begin{subfigure}[b]{0.17\textwidth}
		\includegraphics[width=\textwidth]{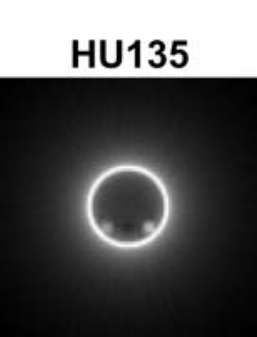}
		\caption{}
	\end{subfigure}
	\caption{(a) Pixel-based attenuation matrix (PAM), (b) Result of applying HU scale to PAM where energy level is 70 keV, Conventional CT (CCT). Primary CT images calculated by utilizing energy intervals where the energy level is (c) 15 keV, (d) 25 keV, (e) 35 keV, (f) 45 keV, (g) 55 keV, (h) 65 keV, (i) 70 keV, (j) 85 keV, (k) 95 keV, (l) 100 keV, and (m) 135 keV} \label{fig:4}
\end{figure}

\begin{figure}[H]
	\centering
	\begin{subfigure}[b]{0.2\textwidth}
		\includegraphics[width=\textwidth]{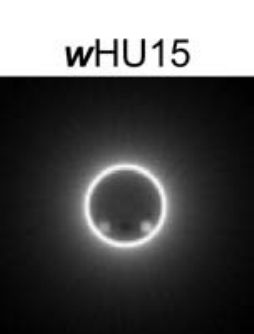}
		\caption{}
	\end{subfigure}
	\hfill
	\begin{subfigure}[b]{0.2\textwidth}
		\includegraphics[width=\textwidth]{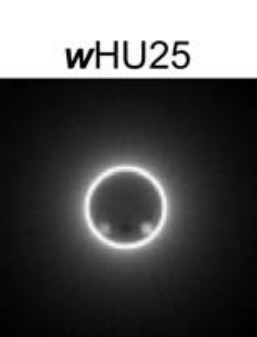}
		\caption{}
	\end{subfigure}
	\hfill
	\begin{subfigure}[b]{0.2\textwidth}
		\includegraphics[width=\textwidth]{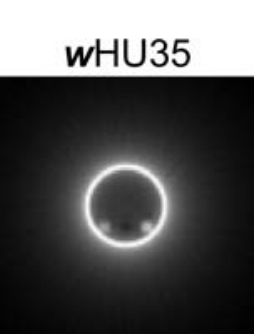}
		\caption{}
	\end{subfigure}
	\hfill
	\begin{subfigure}[b]{0.2\textwidth}
		\includegraphics[width=\textwidth]{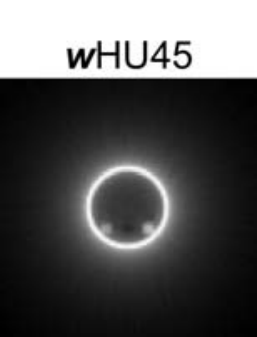}
		\caption{}
	\end{subfigure}
	\hfill
	\begin{subfigure}[b]{0.2\textwidth}
		\includegraphics[width=\textwidth]{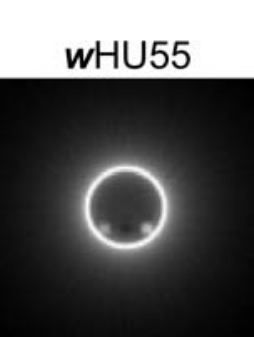}
		\caption{}
	\end{subfigure}
	\hfill
	\begin{subfigure}[b]{0.2\textwidth}
		\includegraphics[width=\textwidth]{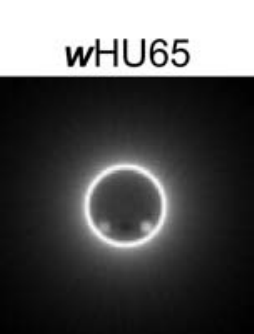}
		\caption{}
	\end{subfigure}
	\hfill
	\begin{subfigure}[b]{0.2\textwidth}
		\includegraphics[width=\textwidth]{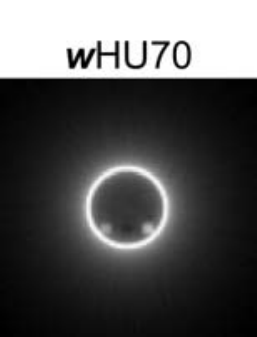}
		\caption{}
	\end{subfigure}
	\hfill
	\begin{subfigure}[b]{0.2\textwidth}
		\includegraphics[width=\textwidth]{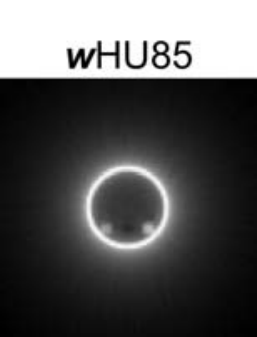}
		\caption{}
	\end{subfigure}
	\hfill
	\begin{subfigure}[b]{0.2\textwidth}
		\includegraphics[width=\textwidth]{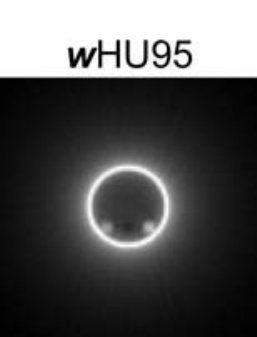}
		\caption{}
	\end{subfigure}
	\hfill
	\begin{subfigure}[b]{0.2\textwidth}
		\includegraphics[width=\textwidth]{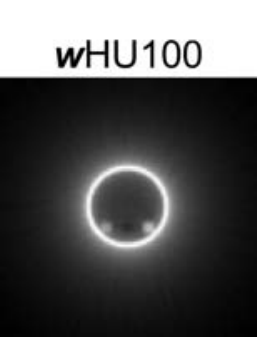}
		\caption{}
	\end{subfigure}
	\hfill
	\begin{subfigure}[b]{0.2\textwidth}
		\includegraphics[width=\textwidth]{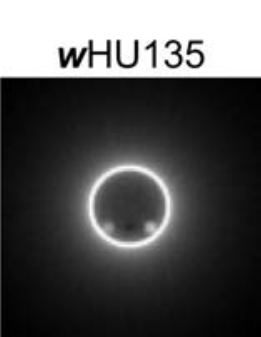}
		\caption{}
	\end{subfigure}
	\caption{Enhanced CT images multiplied by normalized photon flux. 
	Utilizing energy intervals are (a) 15 keV, (b) 25 keV, (c) 35 keV, (d) 45 keV, (e) 55 keV, (f) 65 keV, (g) 70 keV, (h) 85 keV, (i) 95 keV, (j) 100 keV, and (k) 135 keV} \label{fig:5}
\end{figure}

Constructed phantom consists of three rectangular cubes representing skull, rib bone, and lung tissues surrounded by a water cylinder in GATE/GEANT4 environment. Two main reasons for considering these tissues are as follows: 
(1) attenuation coefficients of water are close to the lung, and the same condition exists for the rib bone and skull;
(2) discrimination among tissues is large enough which makes the experiments realistic. 
The radiation source is set to the range of 10-140 keV fan-beam X-ray in a way that could cover double-wedge. Figure~\ref{fig:4} shows primary CT images calculated by utilizing energy intervals in conjunction with the reconstructed CT from the spectrum. Results of enhancing CT images by applying photon flux-oriented weights are shown in Fig.~\ref{fig:5}.

\subsection{Complexity Analysis}

\begin{figure}[H]
	\centering
	\includegraphics[width=0.85\textwidth]{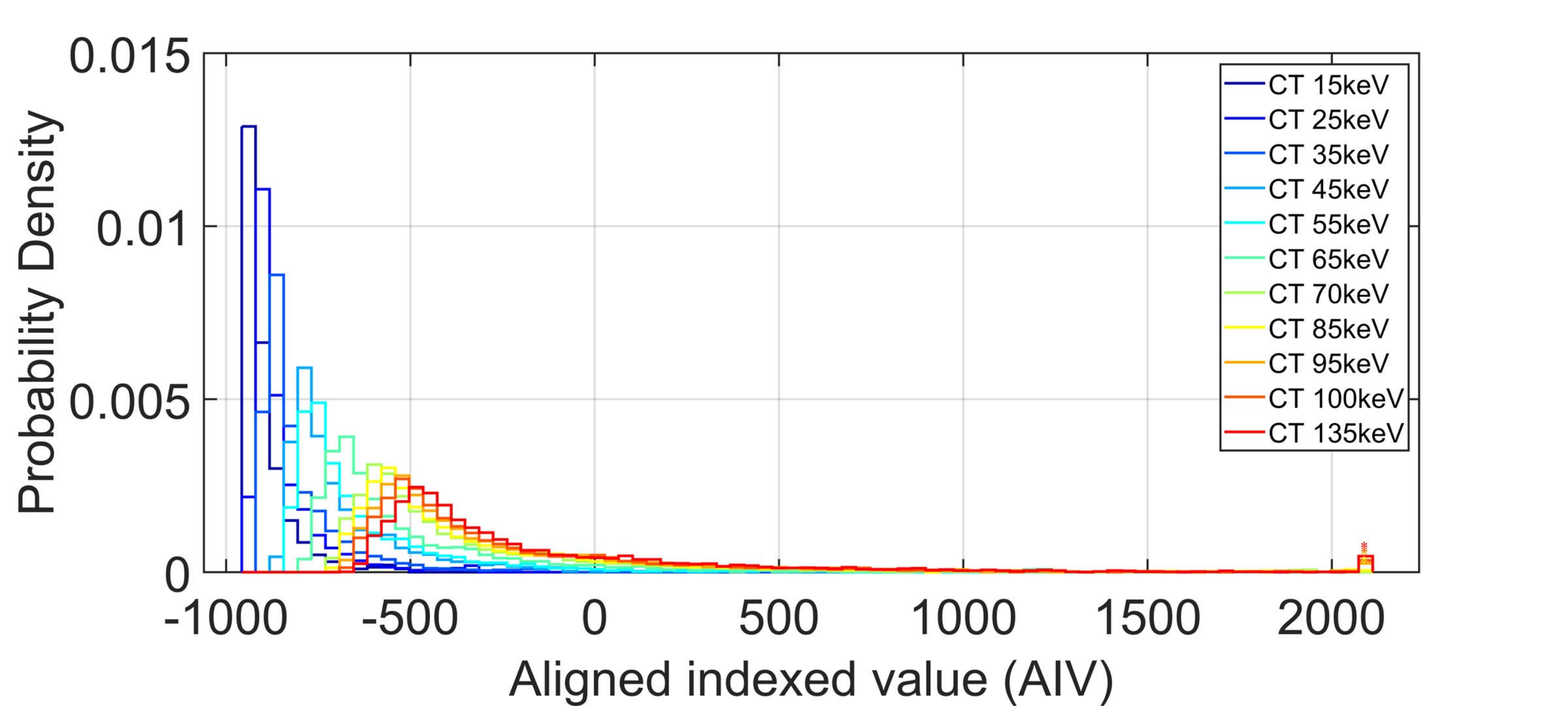}
	\caption{Aligned indexed values (AIV) of all reconstructed CT images from HU scale with respect to the associated probability density functions (PDF).}\label{fig:6}
\end{figure}

Let the simulated CT image be represented by the histogram of indexed values in the range of $I(u, v) \in [-\infty , \infty]$. It contains exactly $K$ entries which are defined by  $h(i) = \mathbf{card}(\{(u,v)|I(u,v) = i\})$. The original dynamics of CT images almost make a quantitative comparison of the associated histograms impossible. As a result, it is reasonable to combine ranges of indexed values into histogram columns to compare the absolute complexity with the relative one. While there is no reference number of bins, grouping data in different bin sizes can reveal different features of the data, see Fig.~\ref{fig:6}, following Scott's normal reference rule \cite{scott1979optimal} (see Eq. \ref{eq:2}).

\begin{equation}
\label{eq:2}
\begin{split}
k = \lceil \frac{\max x- \min x}{h} \rceil, \\
h = \frac{3.5 \sigma}{n^{1/3}},
\end{split}
\end{equation}

where $x$ is a data sample, $\sigma$ is a standard deviation of $x$. The data are obtained as $n$ independent realizations of a bounded probability distribution with smooth density: the histogram remains equally \enquote{rugged} as $n$ tends to infinity. Let $s$ be a width of the distribution, then the frequency of units in a bin is of order $\frac{nh}{s}$ and the relative standard error is of order $\sqrt \frac{s}{nh}$. Comparing to the next bin, the relative change of the frequency is of order $h/s$ provided that the derivative of the density is non-zero. These two are of the same order if $h$ is of order $s/n^{1/3}$, so that $k$ is of order $n^{1/3}$.

For the complexity analysis, a series of numbers $C_{b_{1}}, C_{b_{2}}, \cdots, C_{b_{i}}$ is formed by Eq. \ref{eq:3} to represent the \enquote{temporal} complexity dynamics of the whole image. In terms of complexity, the whole CT spectrum which was calculated at the energy level of 70 keV is considered as the relative complexity. In Eq. \ref{eq:3}, dependence between two quantities is calculated using the correlation coefficient \cite{lee1988thirteen} between two random variables $X$ and $Y$ with expected values $\mu_{X}$ and $\mu_{Y}$ and standard deviations $\sigma_{X}$ and $\mu_{Y}$.

\begin{equation}
\label{eq:3}
C_{b}(CT_{i}, CCT) = \frac{\mathrm{E}[(CT_{i}-\mu_{CT_{i}})(CCT-\mu_{CCT})]}{\sigma_{CT_{i}}\sigma_{CCT}}
\end{equation}

where $E$ is the expected value operator. For CT and modified CT images, we calculate the \enquote{degree of non-constructability} $D(HU_{i})$ and \enquote{generative complexity} $G(wHU_{i})$ measures (see Eq. \ref{eq:4} and \ref{eq:5}). Figure~\ref{fig:7} and \ref{fig:9} show irregular oscillations in the course of the CT and modified CT images when a locally weighted scatterplot smoothing (LOWESS regression). These plots help detect a trend in data that has too much variance resulting in non-significance $p$-values. In this study, we are presented with having to analysis with a smoother regression in which the smoothing factor is set to 0.9. In these plots, the absolute complexity of the reconstructed CT images versus the relative complexity are illustrated to get a handle on quantifying the amount of dispersion.

\begin{equation}
\label{eq:4}
\bigcup_{i \in \mathrm{AIV}} D_{i} := \{D_{i}|D_{i} = \rho_{HU_{i}, CCT}\}
\end{equation}

\begin{equation}
\label{eq:5}
\bigcup_{i \in \mathrm{AIV}} G_{i} := \{G_{i}|G_{i} = \rho_{wHU_{i}, CCT}\}
\end{equation}

\begin{figure}[H]
	\centering
	\begin{subfigure}[b]{0.48\textwidth}
		\includegraphics[width=\textwidth]{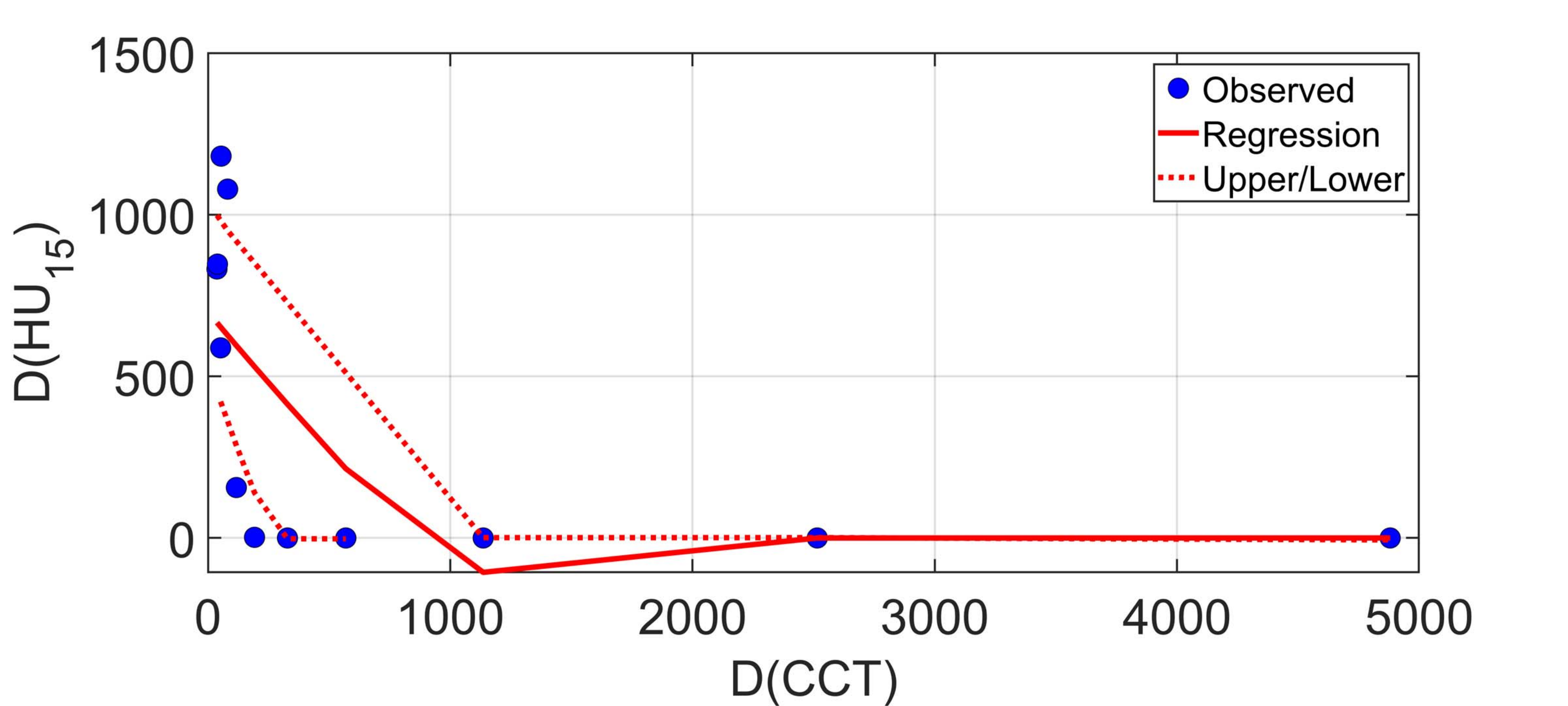}
		\caption{}
	\end{subfigure}
	\hfill
	\begin{subfigure}[b]{0.48\textwidth}
		\includegraphics[width=\textwidth]{Fig_EPS/Fig7a}
		\caption{}
	\end{subfigure}
	\hfill
	\begin{subfigure}[b]{0.48\textwidth}
		\includegraphics[width=\textwidth]{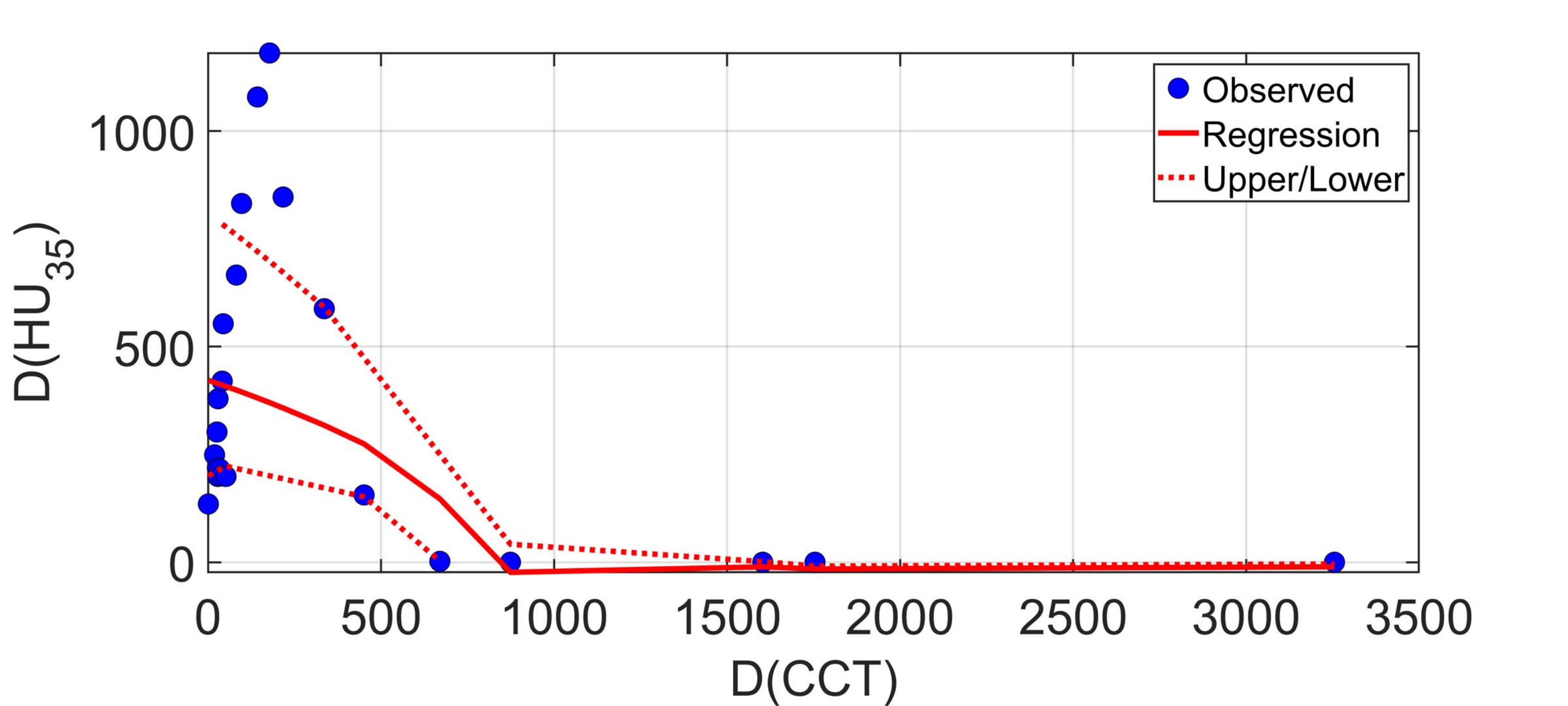}
		\caption{}
	\end{subfigure}
	\hfill
	\begin{subfigure}[b]{0.48\textwidth}
		\includegraphics[width=\textwidth]{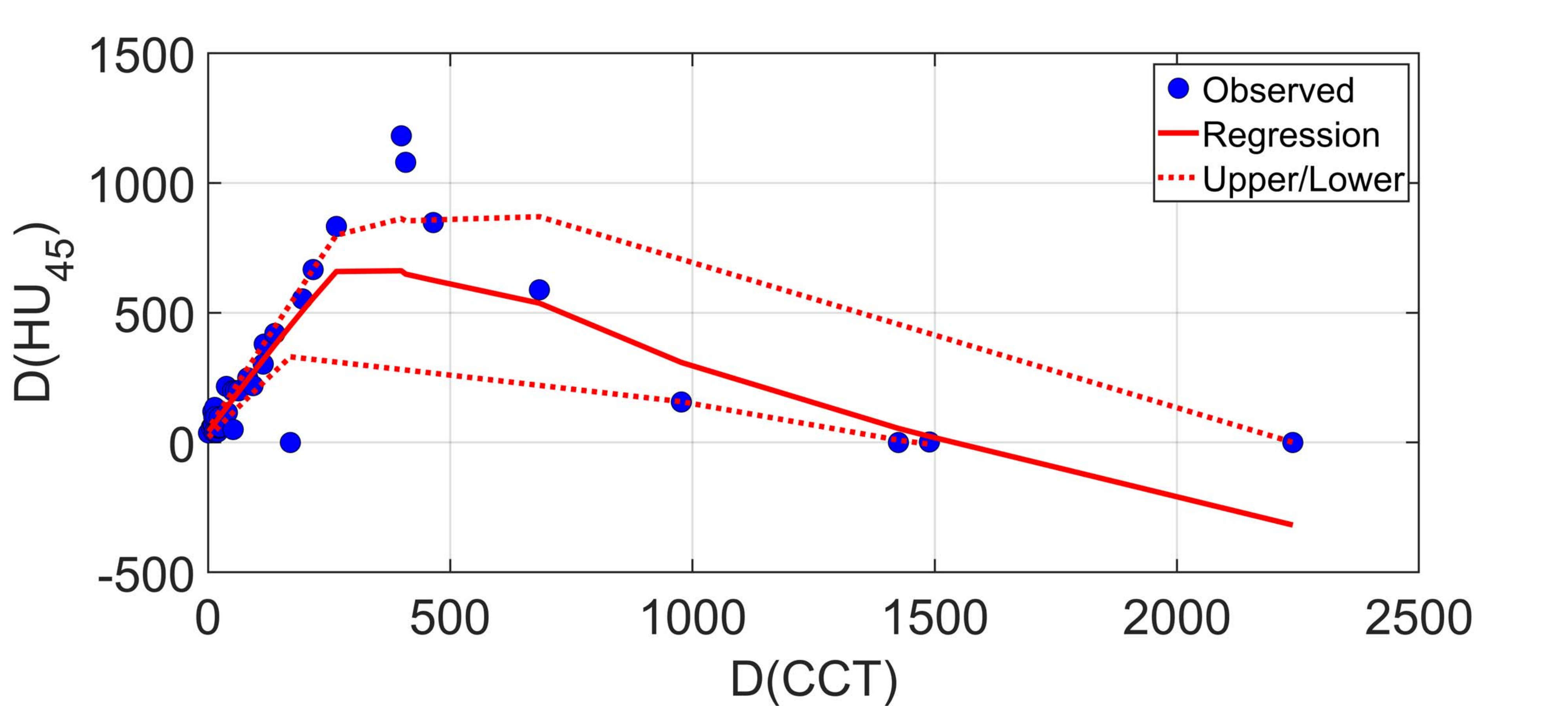}
		\caption{}
	\end{subfigure}
	\hfill
	\begin{subfigure}[b]{0.48\textwidth}
		\includegraphics[width=\textwidth]{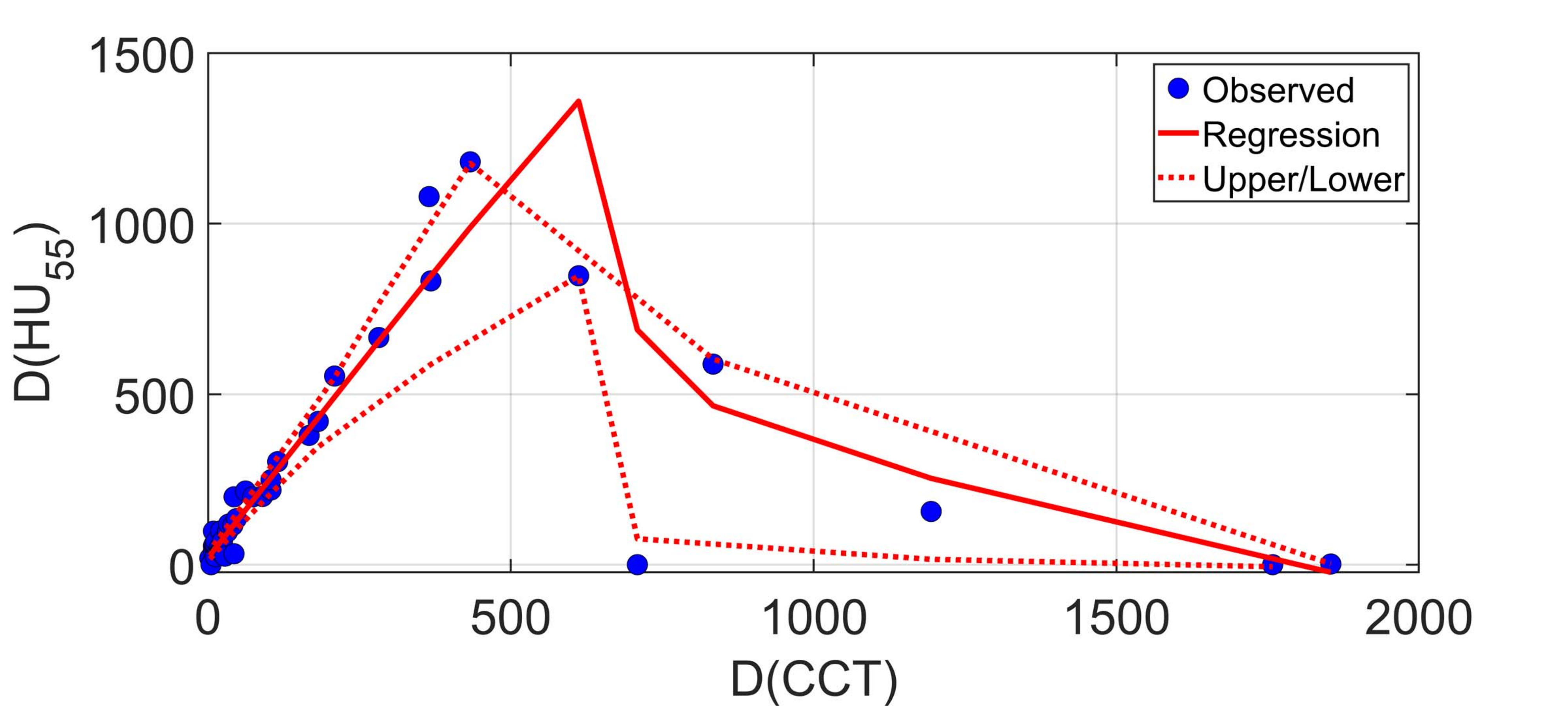}
		\caption{}
	\end{subfigure}
	\hfill
	\begin{subfigure}[b]{0.48\textwidth}
		\includegraphics[width=\textwidth]{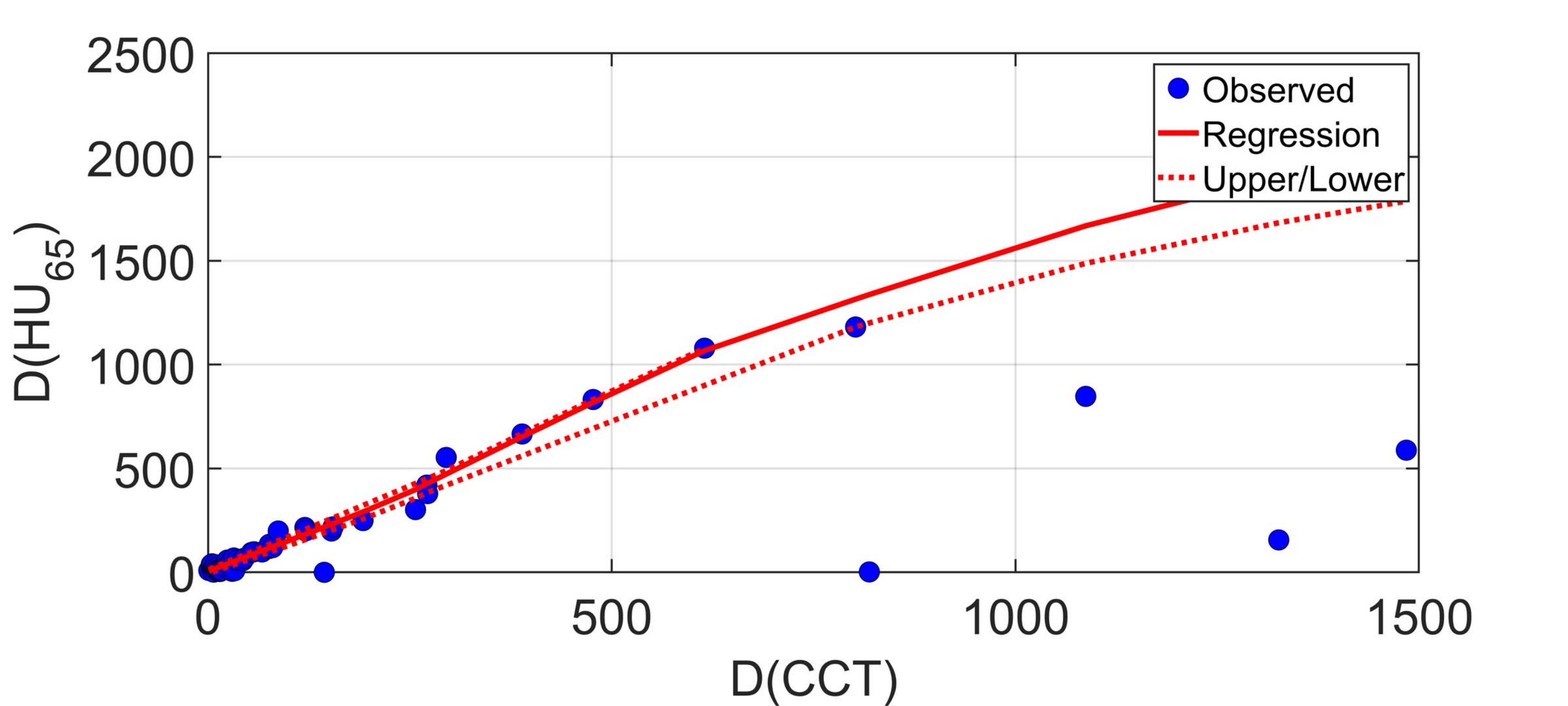}
		\caption{}
	\end{subfigure}
\end{figure}
\begin{figure}[H]\ContinuedFloat
	\begin{subfigure}[b]{0.48\textwidth}
		\includegraphics[width=\textwidth]{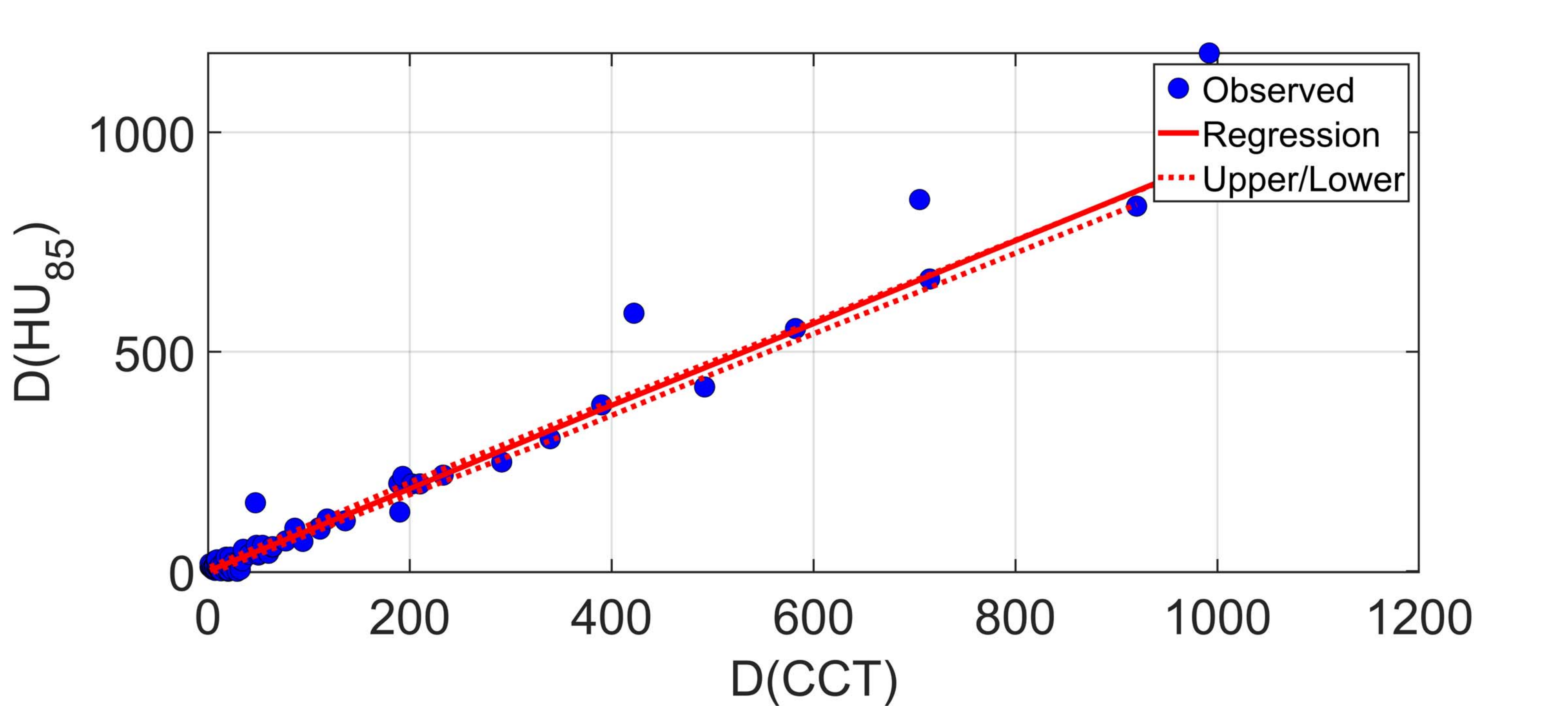}
		\caption{}
	\end{subfigure}
	\hfill
	\begin{subfigure}[b]{0.48\textwidth}
		\includegraphics[width=\textwidth]{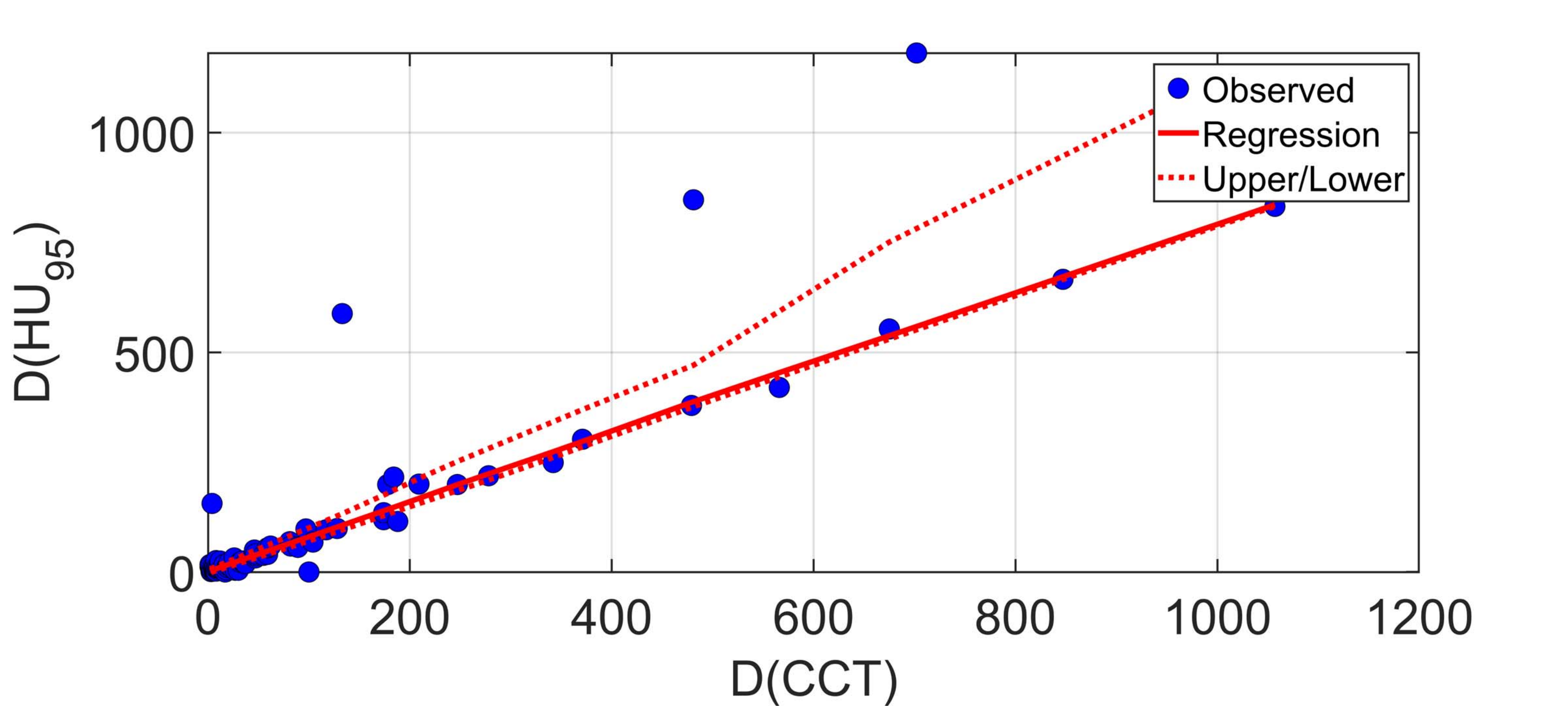}
		\caption{}
	\end{subfigure}
    \hfill	
	\begin{subfigure}[b]{0.48\textwidth}
		\includegraphics[width=\textwidth]{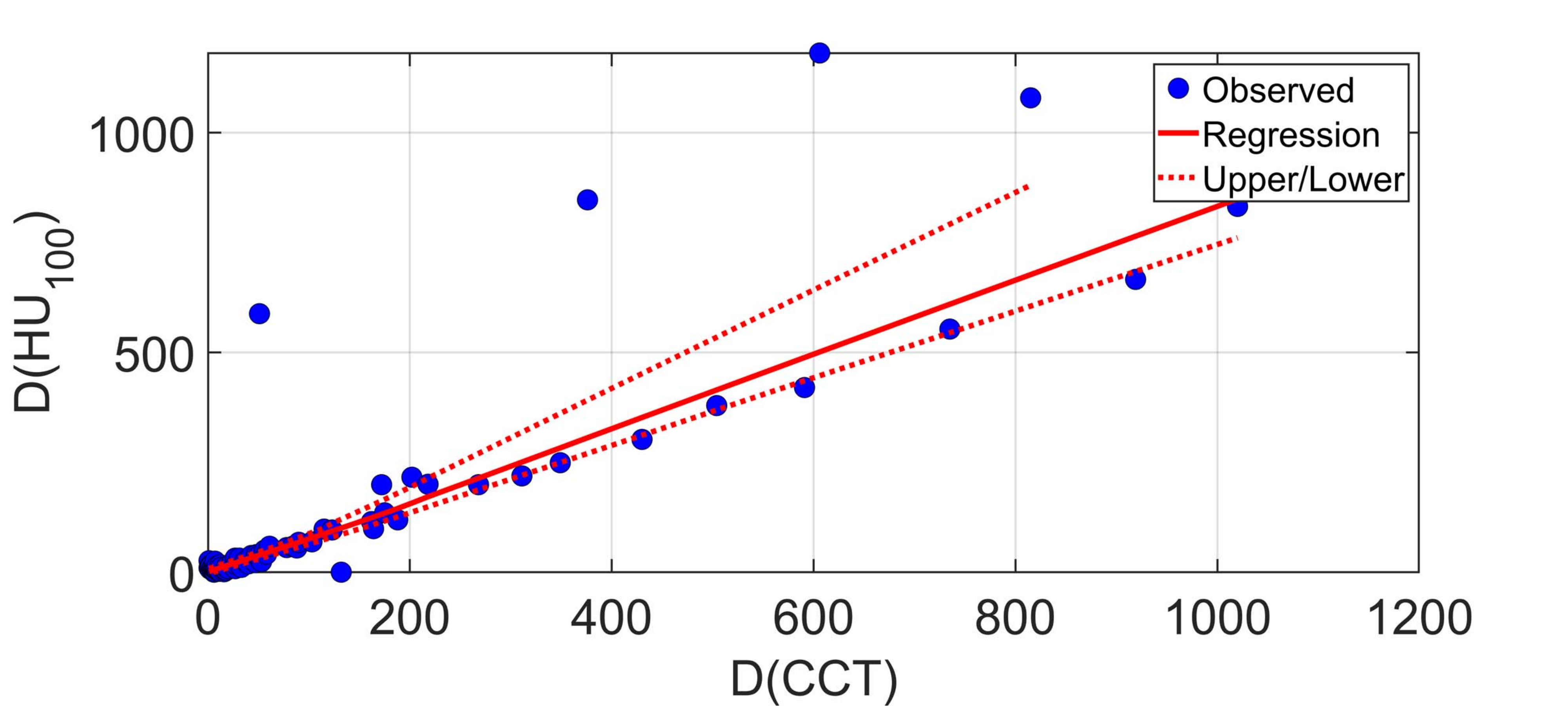}
		\caption{}
	\end{subfigure}
	\hfill
	\begin{subfigure}[b]{0.48\textwidth}
		\includegraphics[width=\textwidth]{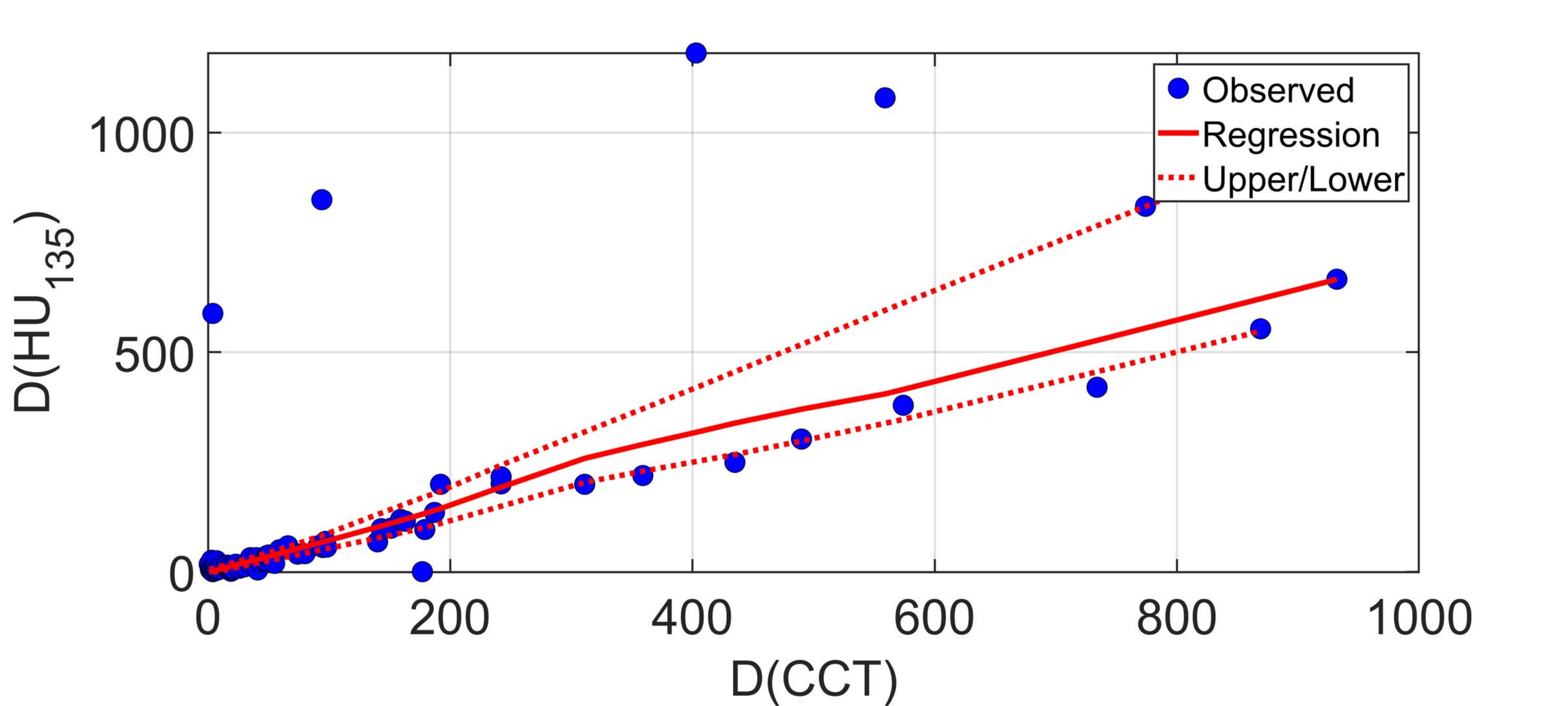}
		\caption{}
	\end{subfigure}
	\caption{Absolute complexity versus relative complexity of the whole bins of the CT images where energy level is (a) 15 keV ($\mu= -893.95, \sigma= 69.65$), (b) 25 keV ($\mu=-845.86, \sigma=101.24$), (c) 35 keV ($\mu=-796.07, \sigma=133.94$), (d) 45 keV ($\mu=-660.11, \sigma=223.24$), (e) 55 keV ($\mu=-621.27, \sigma=248.75$), (f) 65 keV ($\mu=-490.17, \sigma=334.86$), (g) 85 keV ($\mu=-294.91, \sigma=463.10$), (h) 95 keV ($\mu=-238.18, \sigma=500.36$), (i) 100 keV ($\mu=-210.97, \sigma=518.23$), and (j) 135 keV ($\mu=-144.80, \sigma=561.70$)} \label{fig:7}
\end{figure}

The idea of measuring $D$ and $G$ were borrowed from cellular automata theory, where a configuration is called non-constructable if it could not be reached from any other configuration by applying local rules of cell-state transitions \cite{myhill1963converse}, \cite{wolfram1984cellular}, \cite{adamatzky2010generative}. In the context of our model, the degree of non-constructability shows how substantial part of the image cannot be generated from the whole CT spectrum while it must be described by another energy level. The generative complexity $G$, however, shows how difficult it is to generate any particular modified CT images \cite{adamatzky2012diversity}, \cite{ninagawa2014classifying}.

\begin{figure}[H]
	\centering
	\includegraphics[width=0.65\textwidth]{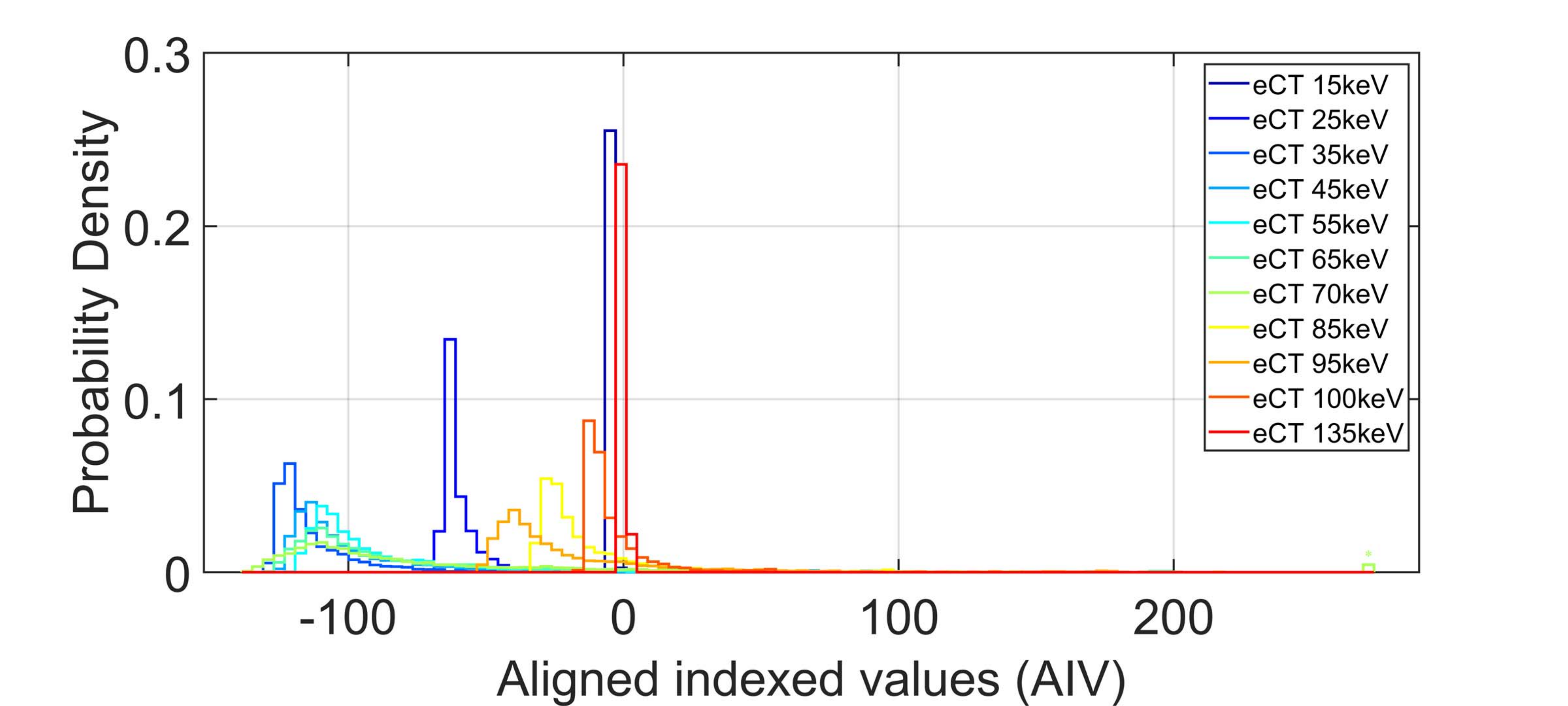}
	\caption{Aligned indexed values (AIV) of all weighted CT images from HU scale with respect to their associated probability density functions (PDF).}\label{fig:8}
\end{figure}

Figure~\ref{fig:7} might reveal that changes in energy levels and dispersion are
analogous to each other. However, higher standard deviation demonstrates that the data points are spread out over a wider range of indexed values. To surpass this diversity, we applied the proposed method to modify CT images which proves that the higher complexity and dispersion are not always analogous to each other (see Fig.~ \ref{fig:8} and \ref{fig:9}). 

A complexity of a system is manifested in its dynamics which may lead to inferring the system as a stochastic one if the structure of the system cannot be recognized. Estimations to Kolmogorov complexity $K$ are used to quantify the randomness degree in CT and enhanced CT images, where they are considered as time series, which are post-processed by different water attenuation coefficients, representing different energy levels. $K(x)$ of an object $x$ is the length, in bits, of the smallest program that, when running on a Universal Turing Machine $U$, produces the object $x$. Although this measure is not computable approximations are possible because $K$ is upper semi-computable meaning that it can be approximated from above. For example, a small size of a lossless compressed version of $x$ is a sufficient test for non-randomness~\cite{ming1990kolmogorov}. 

\begin{figure}[H]
	\centering
	\begin{subfigure}[b]{0.48\textwidth}
		\includegraphics[width=\textwidth]{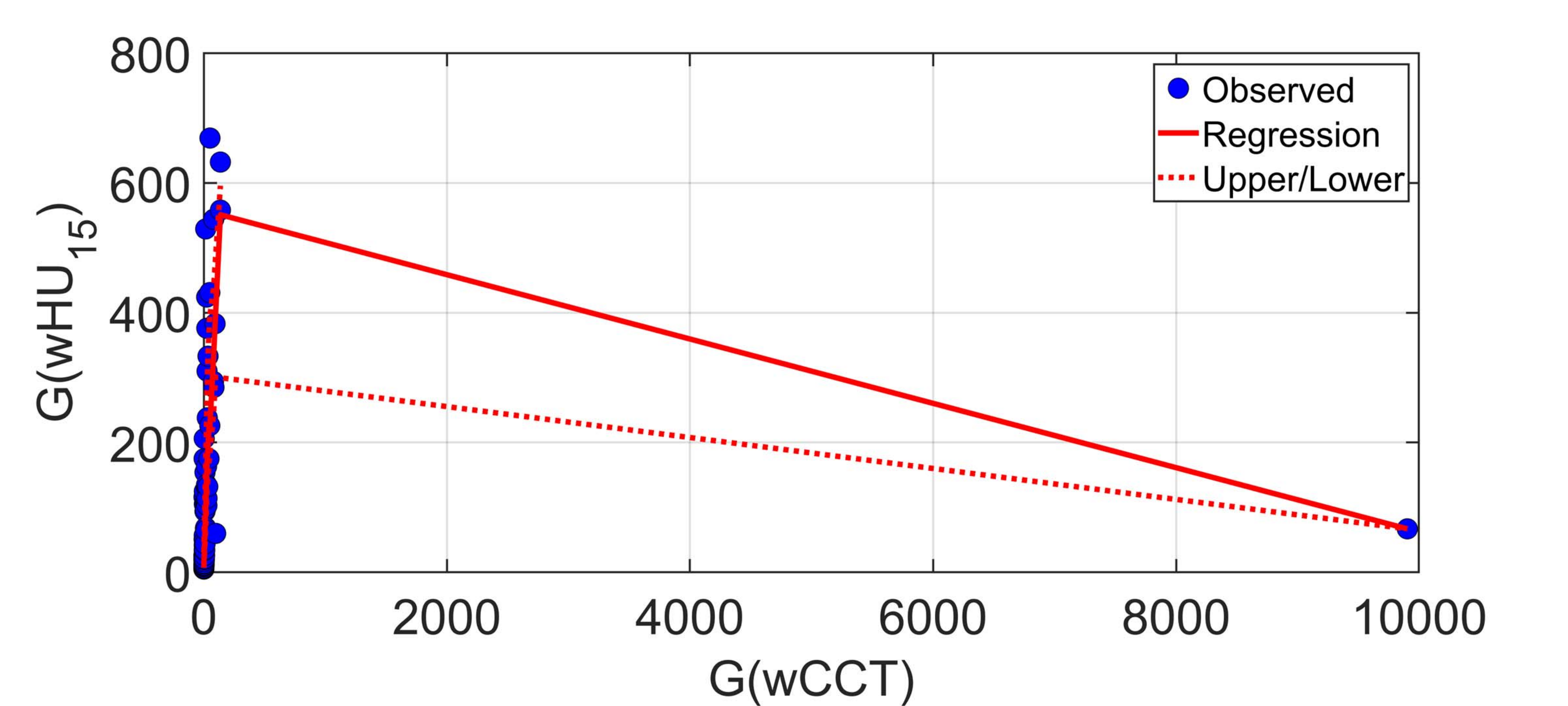}
		\caption{}
	\end{subfigure}
	\begin{subfigure}[b]{0.48\textwidth}
		\includegraphics[width=\textwidth]{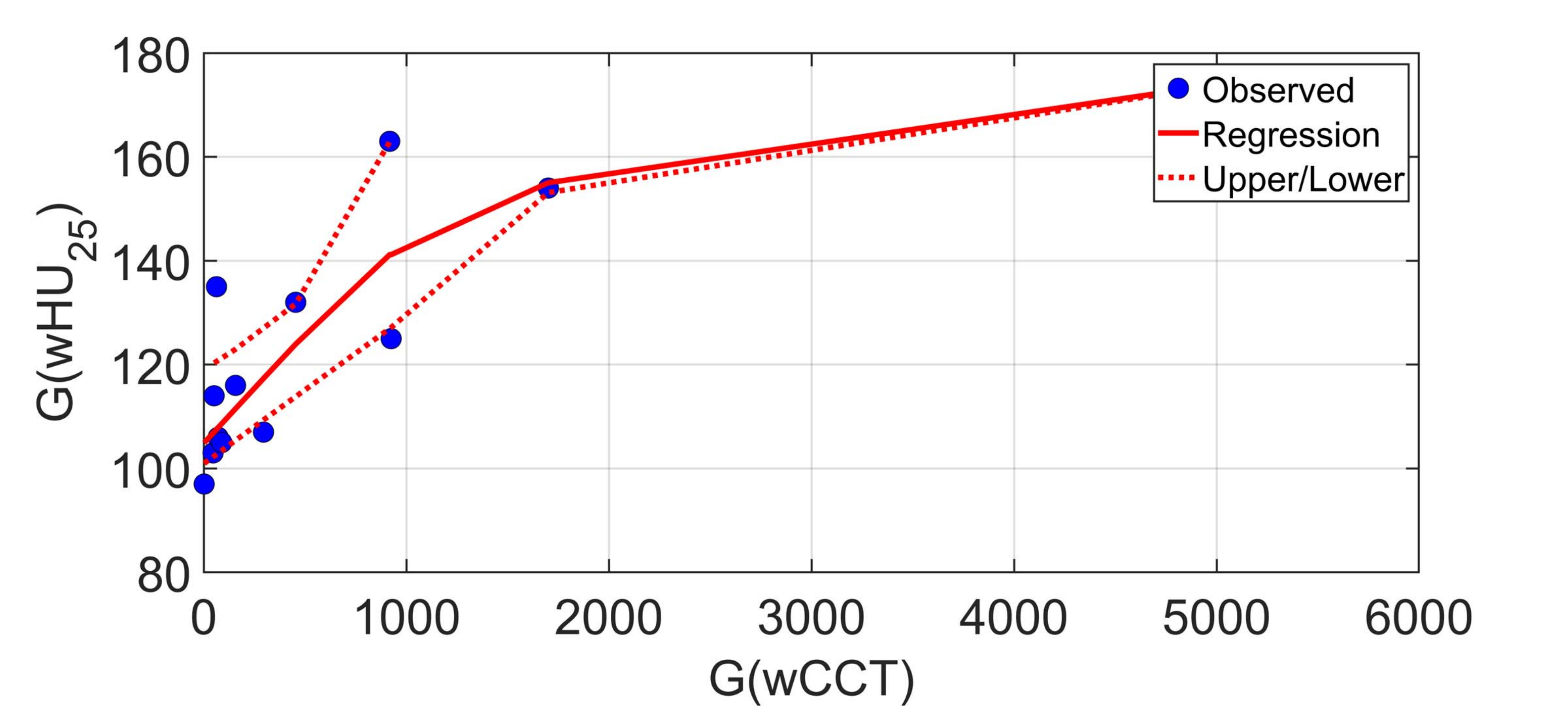}
		\caption{}
	\end{subfigure}
	\hfill
	\begin{subfigure}[b]{0.48\textwidth}
		\includegraphics[width=\textwidth]{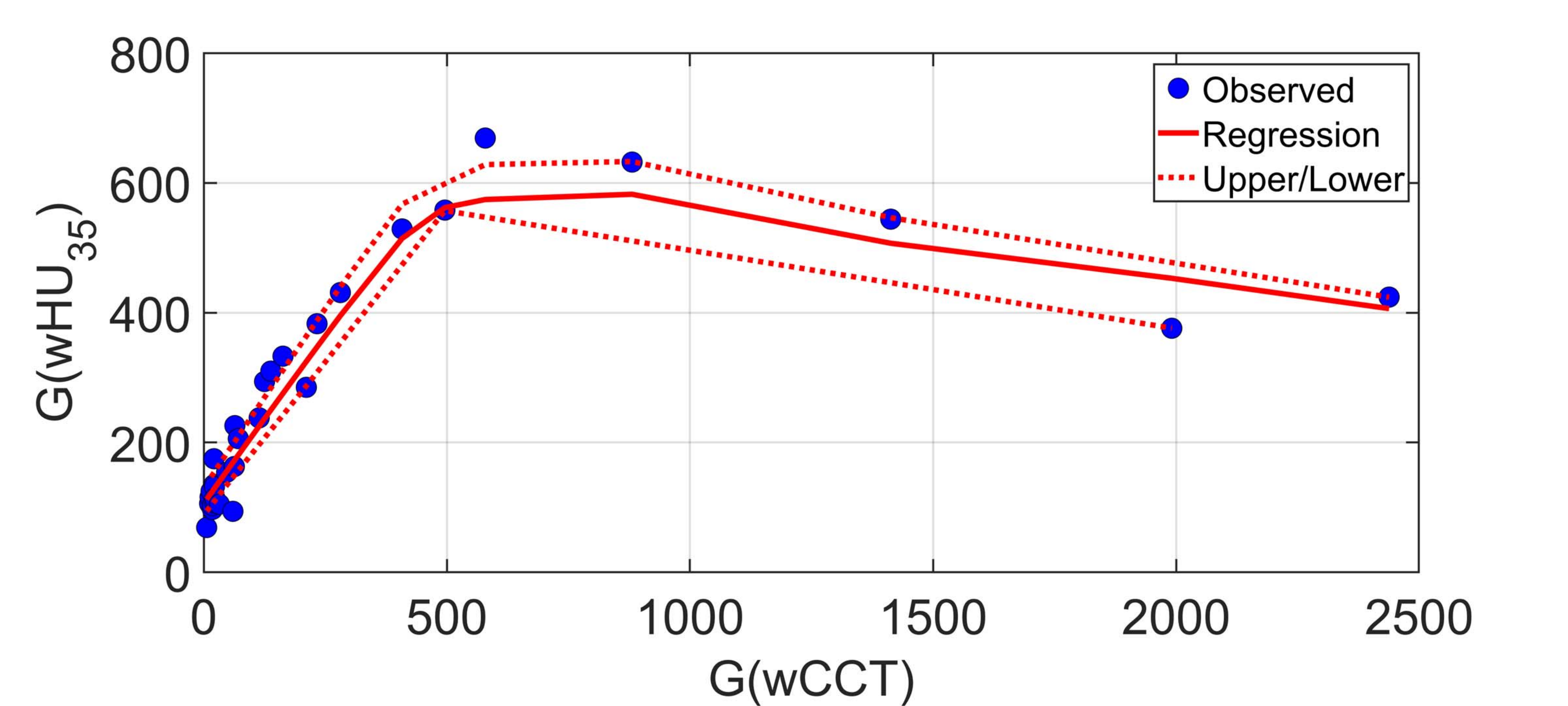}
		\caption{}
	\end{subfigure}
	\hfill
	\begin{subfigure}[b]{0.48\textwidth}
		\includegraphics[width=\textwidth]{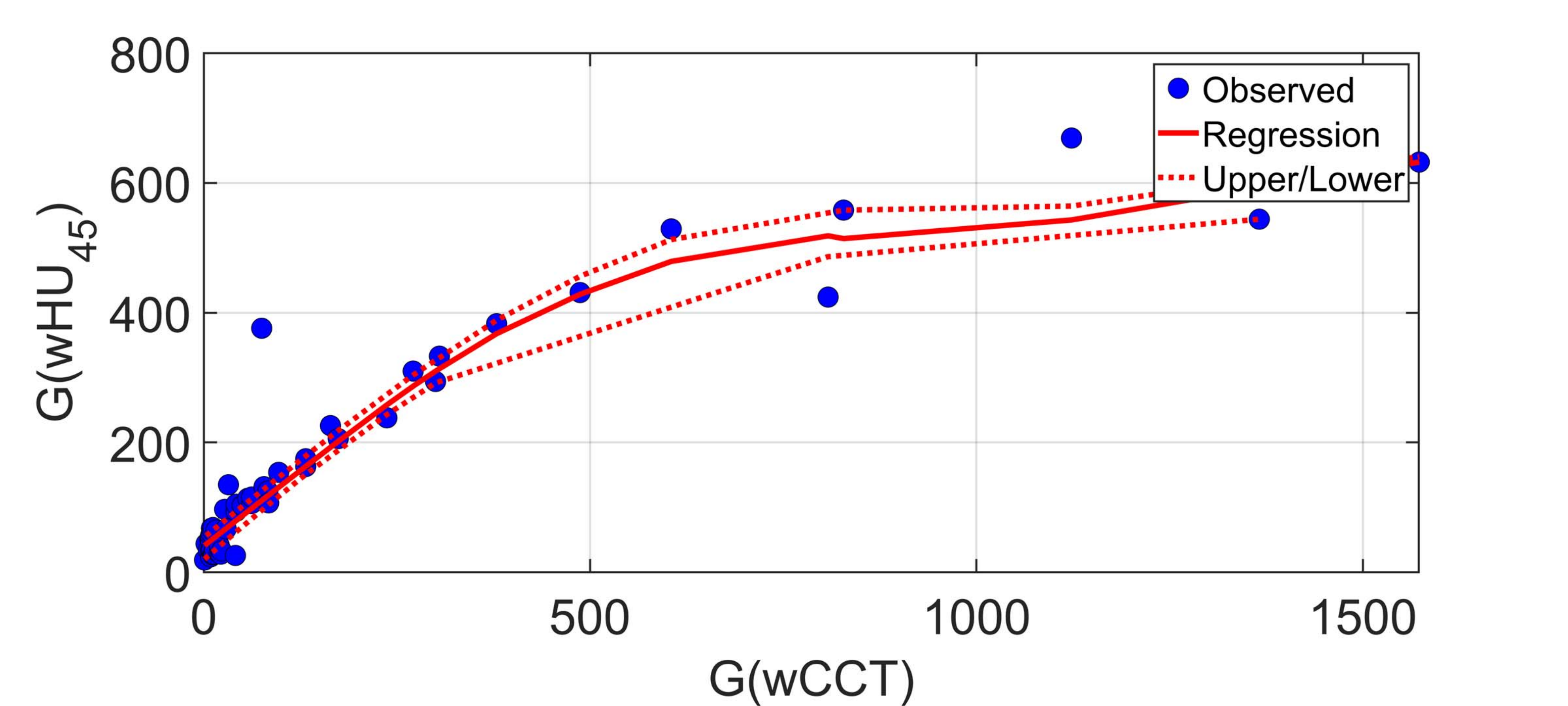}
		\caption{}
	\end{subfigure}
	\hfill
	\begin{subfigure}[b]{0.48\textwidth}
		\includegraphics[width=\textwidth]{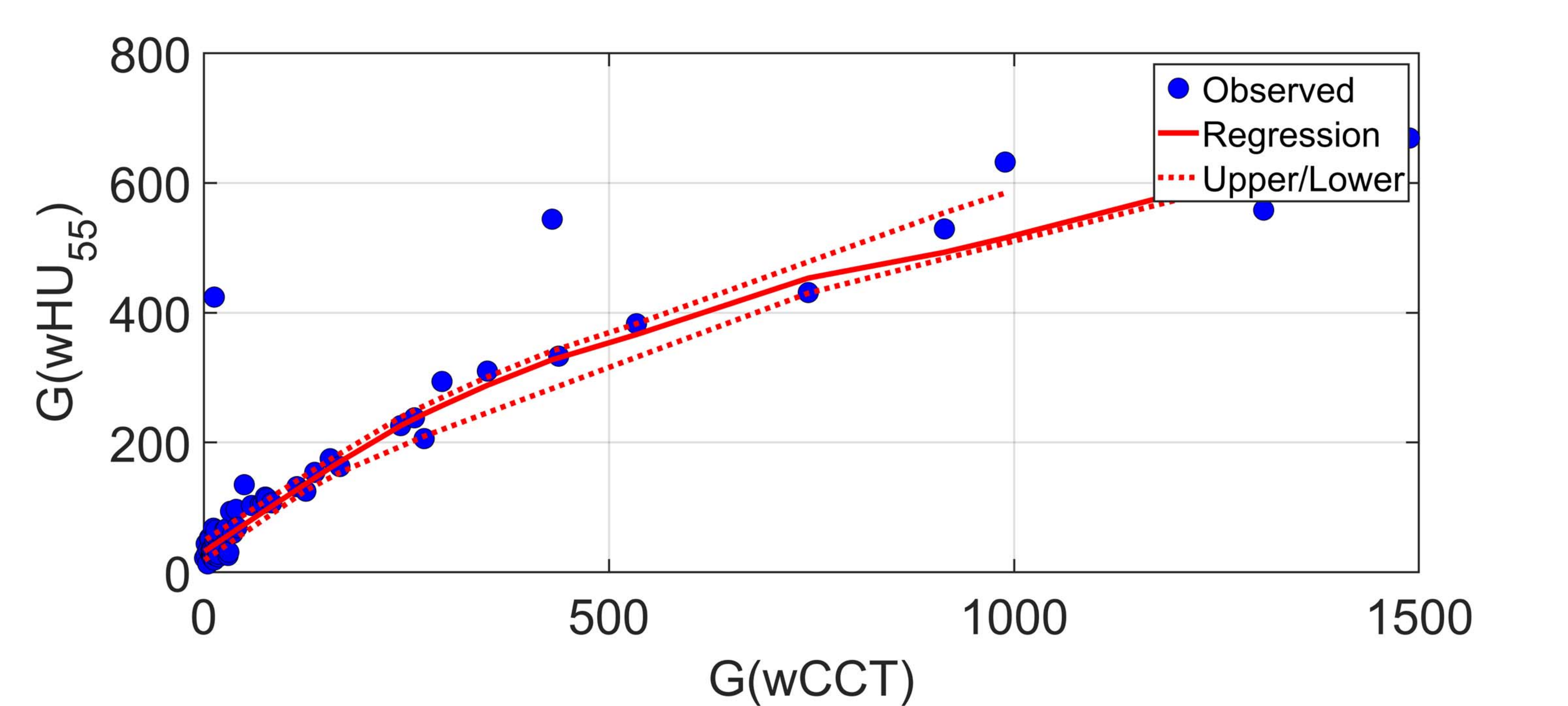}
		\caption{}
	\end{subfigure}
	\hfill
	\begin{subfigure}[b]{0.48\textwidth}
		\includegraphics[width=\textwidth]{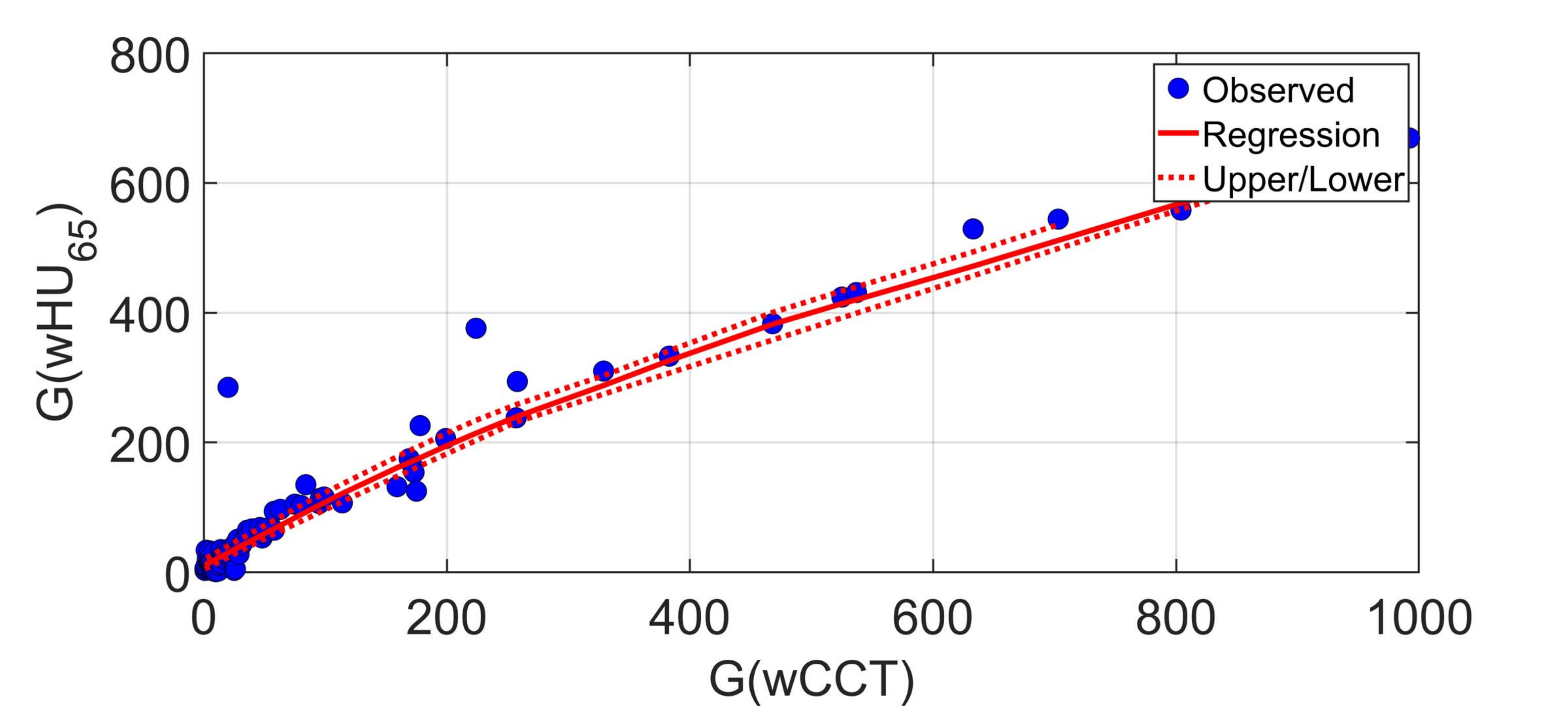}
		\caption{}
	\end{subfigure}
	\hfill
	\begin{subfigure}[b]{0.48\textwidth}
		\includegraphics[width=\textwidth]{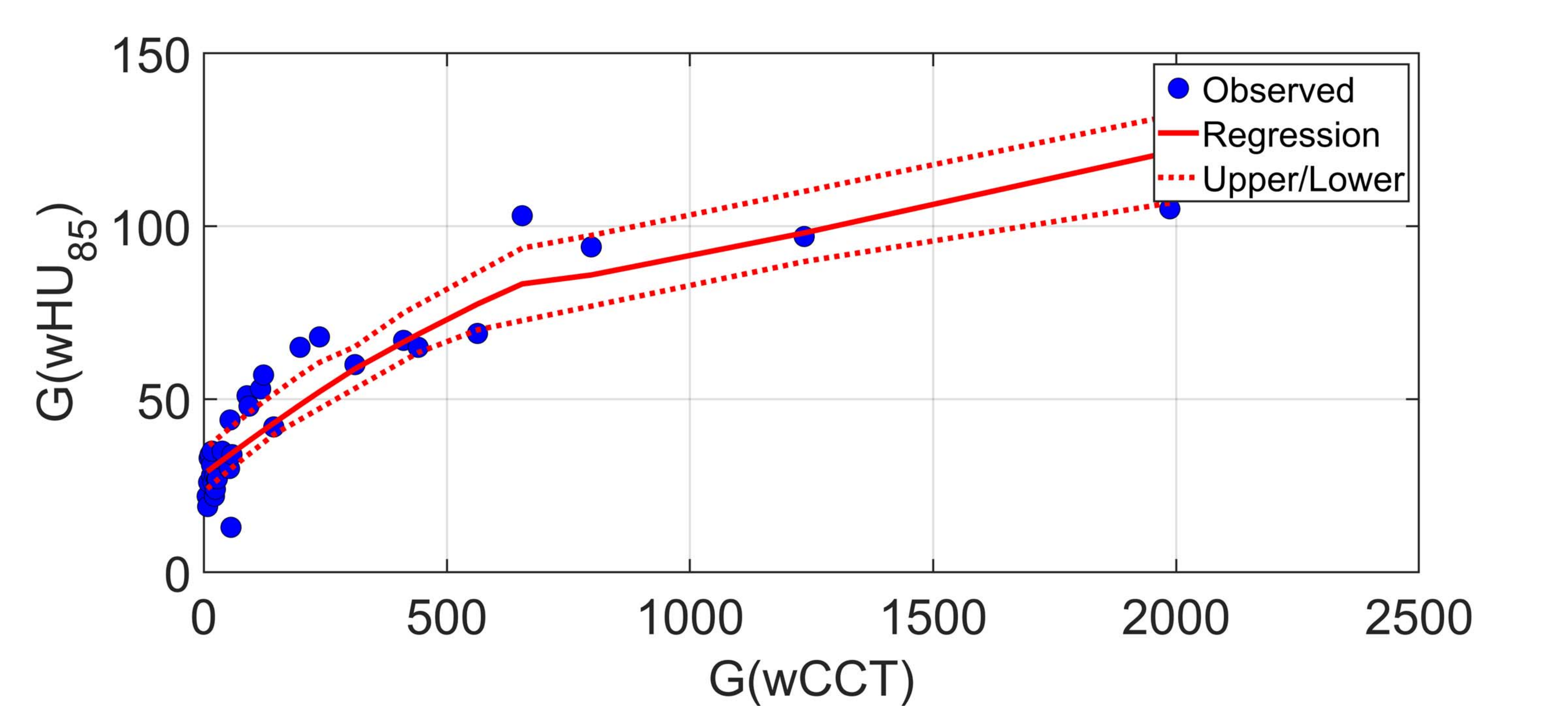}
		\caption{}
	\end{subfigure}
	\hfill
	\begin{subfigure}[b]{0.48\textwidth}
		\includegraphics[width=\textwidth]{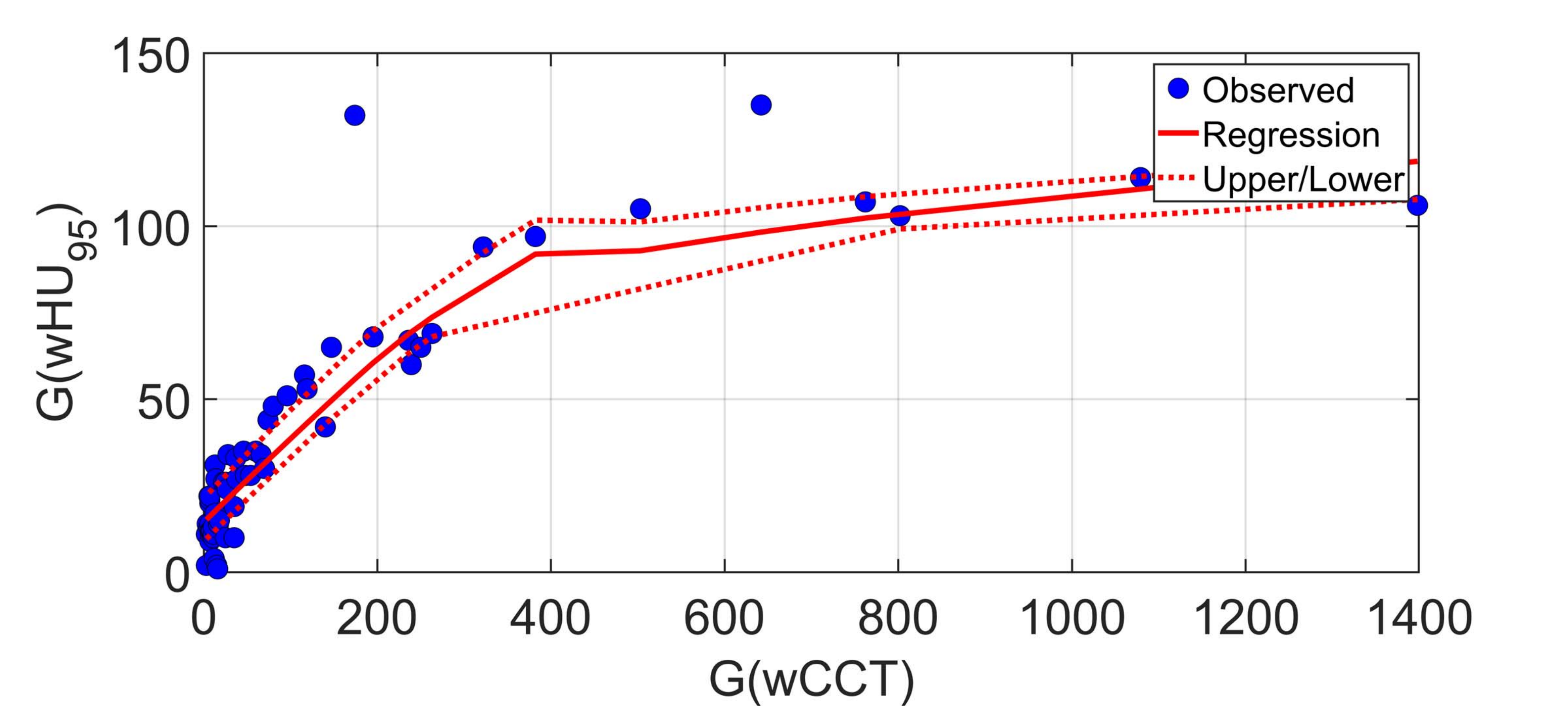}
		\caption{}
	\end{subfigure}
\end{figure}
\begin{figure}[H]\ContinuedFloat
	\begin{subfigure}[b]{0.48\textwidth}
		\includegraphics[width=\textwidth]{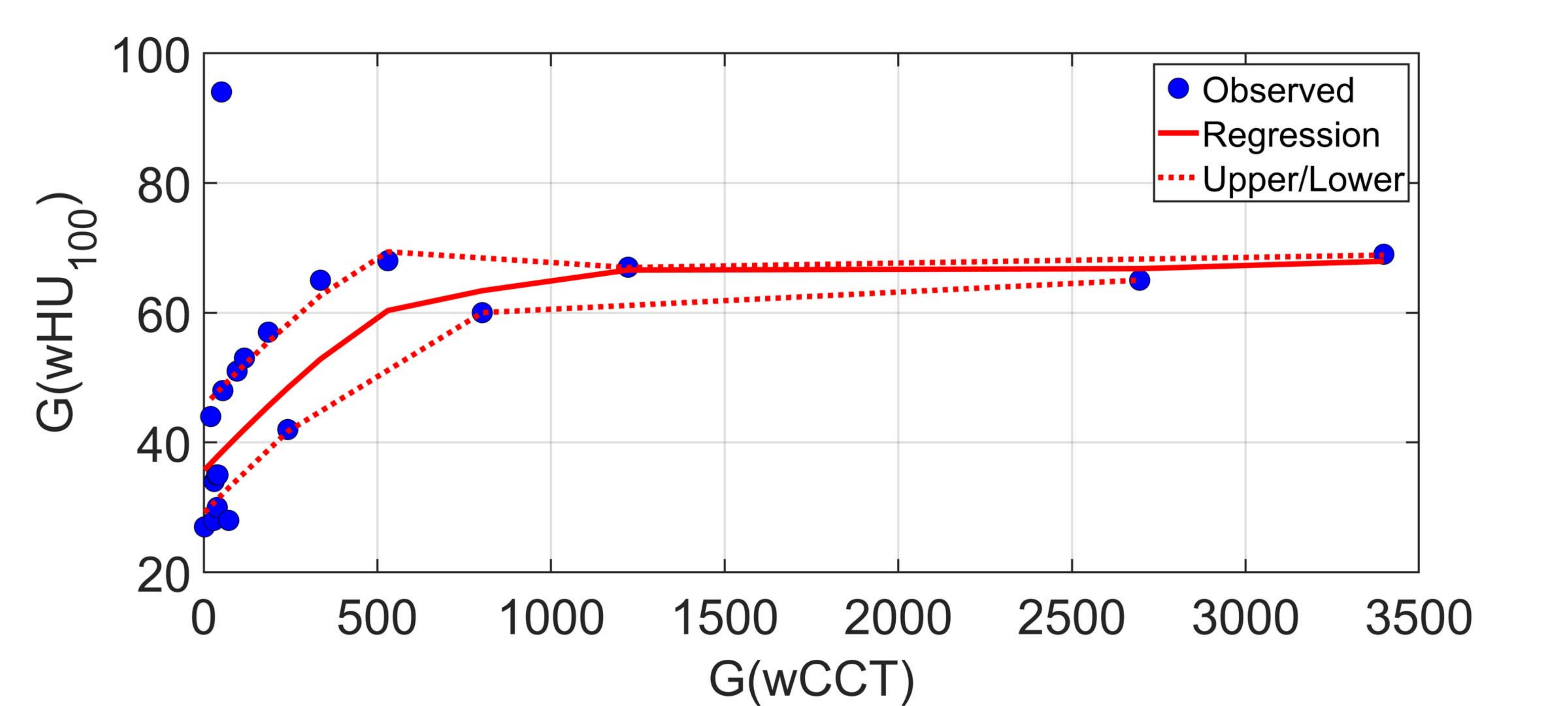}
		\caption{}
	\end{subfigure}
	\hfill
	\begin{subfigure}[b]{0.48\textwidth}
		\includegraphics[width=\textwidth]{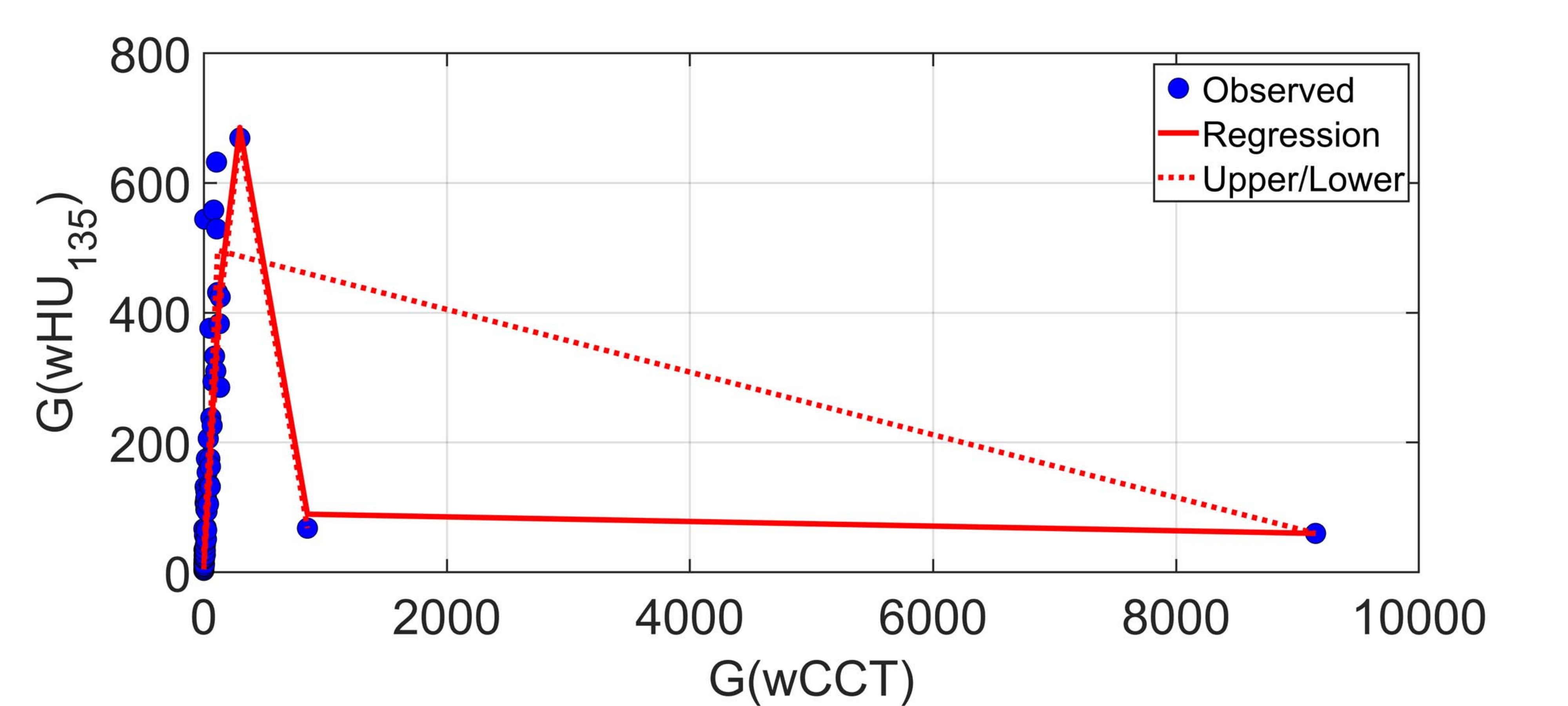}
		\caption{}
	\end{subfigure}
	\caption{Absolute complexity versus relative complexity of the modified CT images where the energy level is (a) 15 keV ($\mu= -893.95, \sigma= 69.65$), (b) 25 keV ($\mu=-845.86, \sigma=101.24$), (c) 35 keV ($\mu=-796.07, \sigma=133.94$), (d) 45 keV ($\mu=-660.11, \sigma=223.24$), (e) 55 keV ($\mu=-621.27, \sigma=248.75$), (f) 65 keV ($\mu=-490.17, \sigma=334.86$), (g) 85 keV ($\mu=-294.91, \sigma=463.10$), (h) 95 keV ($\mu=-238.18, \sigma=500.36$), (i) 100 keV ($\mu=-210.97, \sigma=518.23$), and (j) 135 keV ($\mu=-144.80, \sigma=561.70$)} \label{fig:9}
\end{figure}

Estimations to Kolmogorov complexity of a time series $\{x_{i}\}, i = 1, 2, 3, 4, \cdots, N$ by the LZW algorithm can be carried out as is in Eq. \ref{eq:6}:

\begin{gather}
\label{eq:6}
s(i) = \begin{cases}0, & x_{i} < x_{*} \\1, & x_{i}  \geq  x_{*}\end{cases}, \\
c(N) = O(b(N)), \quad b(N) = \frac{N}{\log_{2}N}, \\
C_{k}=\frac{c(N)}{b(N)}=c(N)\frac{\log_{2}N}{N}.
\end{gather}

where $x_{*}$ is the mean value of the time series to be the threshold, $c(N)$ is the minimum number of distinct patterns contained in a given character sequence, and $C_{k}(N)$ represents the information quantity of a time series to demonstrate if it a periodic or random time series. For a nonlinear time series, $C_{k}(N)$ varies between 0 and 1,
although Hu et al. \cite{hu2006analysis} have demonstrated that $C_{k}$ can be larger than 1. 

However, popular lossless compression algorithms such as those based in LZW are closer to entropy than to $K$~\cite{emergence} and thus alternatives have been introduced. Methods designed and tested to outperform compression algorithms have been introduced~\cite{zenil2016decomposition} and are based on approximations to algorithmic probability as it is deeply connected to $K$. Algorithmic probability is the probability of an object $x$ to be produced by a Universal Turing Machine and according to the algorithmic Coding Theorem it is inversely proportional to $K$ and can be empirically estimated from e.g. the output frequency of small Turing machines by the so-called Coding Theorem Method~\cite{soler2014calculating} (CTM) and the aggregation of these values via an algorithm called the Block Decomposition Method (BDM).

\begin{figure}[H]
    \centering
    \begin{subfigure}[b]{\textwidth}
    \centering
    \includegraphics[width=0.85\linewidth]{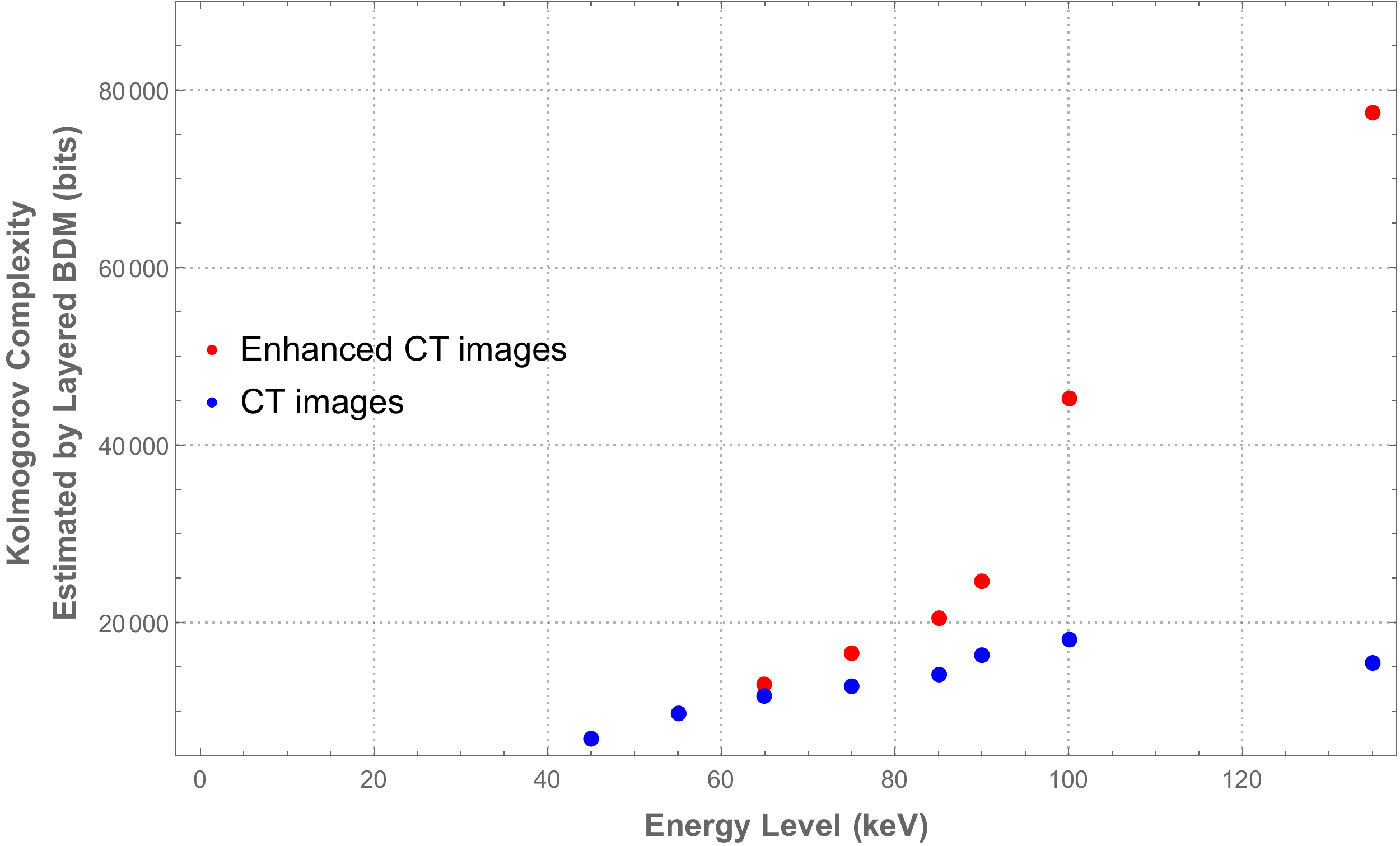}
    \caption{Kolmogorov complexity estimated by the Layered Block Decomposition Method in CT images and enhanced CT images, Spearman $\rho = 0.972$, $\textit{p-value} = 5.58 \times 10^-{7}$ }
    \label{fig:layeredBDM}
    \end{subfigure}
    
     \begin{subfigure}[b]{\textwidth}
     \centering
    \includegraphics[width=0.85\linewidth]{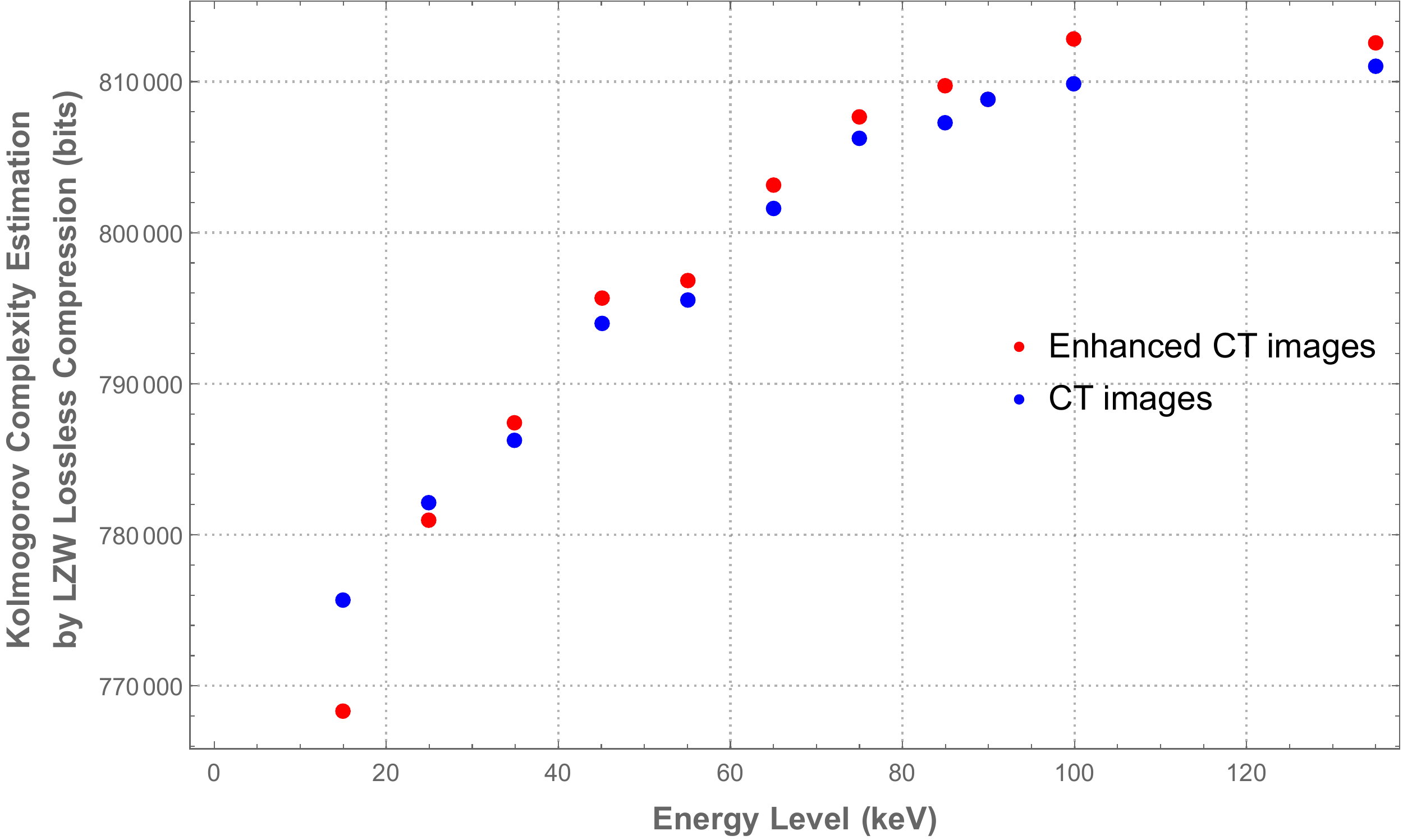}
    \caption{Kolmogorov complexity estimated by Lempel-Ziv-Welch (LZW) lossless compression length in CT images and enhanced CT images, Spearman $\rho = 0.98$, $\textit{p-value} = 8.4 \times 10^-{8}$ }
    \label{fig:lempelziv}
    \end{subfigure}
    \caption{Estimations of Kolmogorov complexity}
    \label{fig:kolmoplot}
\end{figure}

Figure~\ref{fig:layeredBDM} shows the estimations to $K$ for CT and enhanced CT images using the Layered Block Decomposition Method, a variant of BDM for grayscale and multichannel images~\cite{Rueda-Toicen2018code, Rueda-Toicen2018paper}. In Layered BDM, images are quantized and binarized in $q$ digital levels before aggregating known CTM values for the blocks in which each layer is decomposed. The coarse graining of the Kolmogorov complexity estimation is defined by the number of digital levels in which an image is quantized, e.g. 256 levels (int-8 quantization), 65536 levels (int-16 quantization), 4294967296 levels (int-32 quantization), etc. In the results shown on Figure~\ref{fig:layeredBDM} $q = 256$, as the images where quantized and binarized on the range $(0, 255)$.

\begin{figure}[H]
    \centering
    \includegraphics[width=\linewidth]{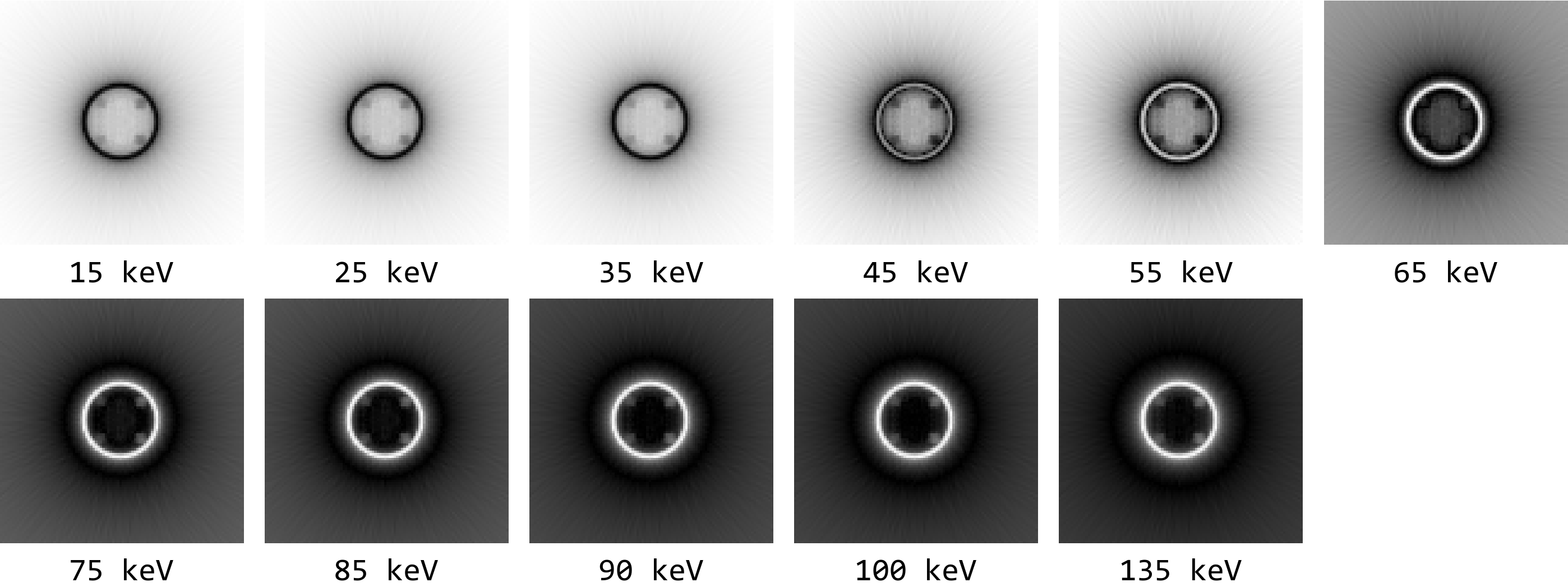}
    \caption{Differences between the enhanced and non-enhanced versions of the CT images at the studied energy levels, lighter areas represent values with smaller pixel-wise differences.}
    \label{fig:image_differences}
\end{figure}

\begin{algorithm}[H]
\SetAlgoLined
\SetKwProg{Fn}{Function}{ is }{end}
\DontPrintSemicolon
\SetAlgoLined
\emph{// CTMs is a hashtable with binary 2D blocks as keys,}\;
\emph{// their respective values being estimations of Kolmogorov complexity}\;
\emph{// obtained through the Coding Theorem Method}\;

\BlankLine

\Fn{LayeredBDM(grayImage, CTMs, blockSize, blockOffset, q)}{

\emph{// the image is quantized in q digital levels}\;

grayImage $\longleftarrow$  quantize(grayImage, q) 

blocksList $\longleftarrow \{ \}$

\For{ $i$ in 1 \KwTo $q$}{

\emph{// the quantized image is binarized in q digital layers}\;

    binImage $\longleftarrow$  binarize(grayImage, q)\;
    blocks $\leftarrow$ partition (binImage, blockSize, blockOffset)\;
    blocksList.append(blocks)\;
}
\BlankLine
\emph{// we count the appearance of all binary blocks through all layers}\;
\emph{// and store the count of each into a hash table with the blocks as keys}\;
\emph{// and the block counts as values }\;
blockHT(blocks:blockCount) $\longleftarrow$ countBlocks(blockList)\;
\BlankLine
\emph{// the blocks' CTM values are retrieved from the CTMs hashtable}\;
\emph{// these and the $\log_{2}$ of the cardinality of each are added} \;
\textit{l-BDM} $\leftarrow$ \textit{CTMs(keys(blockHT))} $+$ $\log_{2}$\textit{(values(blockHT))}\;
\BlankLine
	\KwRet{l-BDM}

} 
\caption{Layered Block Decomposition Method for grayscale images}
\label{layeredBDMalgorithm}
\end{algorithm}

Figure~\ref{fig:kolmoplot} shows the estimated Kolmogorov complexities (KC) of CT and enhanced CT images. Fig.~\ref{fig:layeredBDM} shows the estimated Kolmogorov complexity obtained through the Layered Block Decomposition method\footnote{code and analysis available at: \url{https://github.com/andandandand/ImageAnalysisWithAlgorithmicInformation}}~\cite{Rueda-Toicen2018code, Rueda-Toicen2018paper, zenil2016decomposition, soler2014calculating, zenil2014correlation}, described in Fig.~\ref{layeredBDMalgorithm}, and Fig.~\ref{fig:lempelziv} shows the KC estimation obtained by the lossless compression algorithm Lempel-Ziv-Welch as implemented in the Wolfram Language's \texttt{Compress} function~\cite{compress}.
Both Fig.~\ref{fig:layeredBDM}, and Fig.~\ref{fig:lempelziv} show an almost monotonic increase increase in Kolmogorov complexity when the energy level increases. In Fig.~\ref{fig:layeredBDM} we appreciate a small difference in KC between enhanced and non-enhanced CT when the energy levels are below 65 keV. The KC differences in bits between enhanced and non-enhanced data increases more when the energy levels goes up in Fig.~\ref{fig:layeredBDM} than in Fig.~\ref{fig:lempelziv}. The KC estimations in bits obtained with layered BDM are an order of magnitude below the ones obtained with lossless compression length. A Spearman rank correlation test between the KC values obtained with layered BDM and compression length in CT data gives $\rho = 0.96$ with p-value $= 1.91 \times 10^{-6}$. In the enhanced CT data, the Spearman rank test gives $\rho = 0.97$ with p-value $= 5.32 \times 10^{-7} $. Visual inspection of the pixel-wise differences between enhanced and non-enhanced versions of the images, shown in Fig.~\ref{fig:image_differences}, indicates that the characterizations obtained by Layered BDM are more sensitive to morphological changes in the images than the ones obtained with lossless compression. 

Entropy quantifies the unpredictability of a state, which shows its average information content. Because of its crucial dependency on the probabilistic model, it is not a universal measure of complexity. Indeed, entropy quantifies these considerations when a probability distribution of the source data is known \cite{shannon1948mathematical}. The benefit of utilizing entropy in the context of complexity is that it only considers the probability of observing a specific event, so it does not express any interpretation of the meaning of the events themselves. In this study, we calculate the following entropies: (1) Approximate Entropy, (2) Conditional Entropy, (3) Corrected Conditional Entropy, (4) Sample Entropy, (5) Fuzzy Entropy, and (6) Permutational Entropy, each of which could reveal a part of associated complexity to the CT data.

\begin{figure}[H]
	\centering
	\includegraphics[width=1\textwidth]{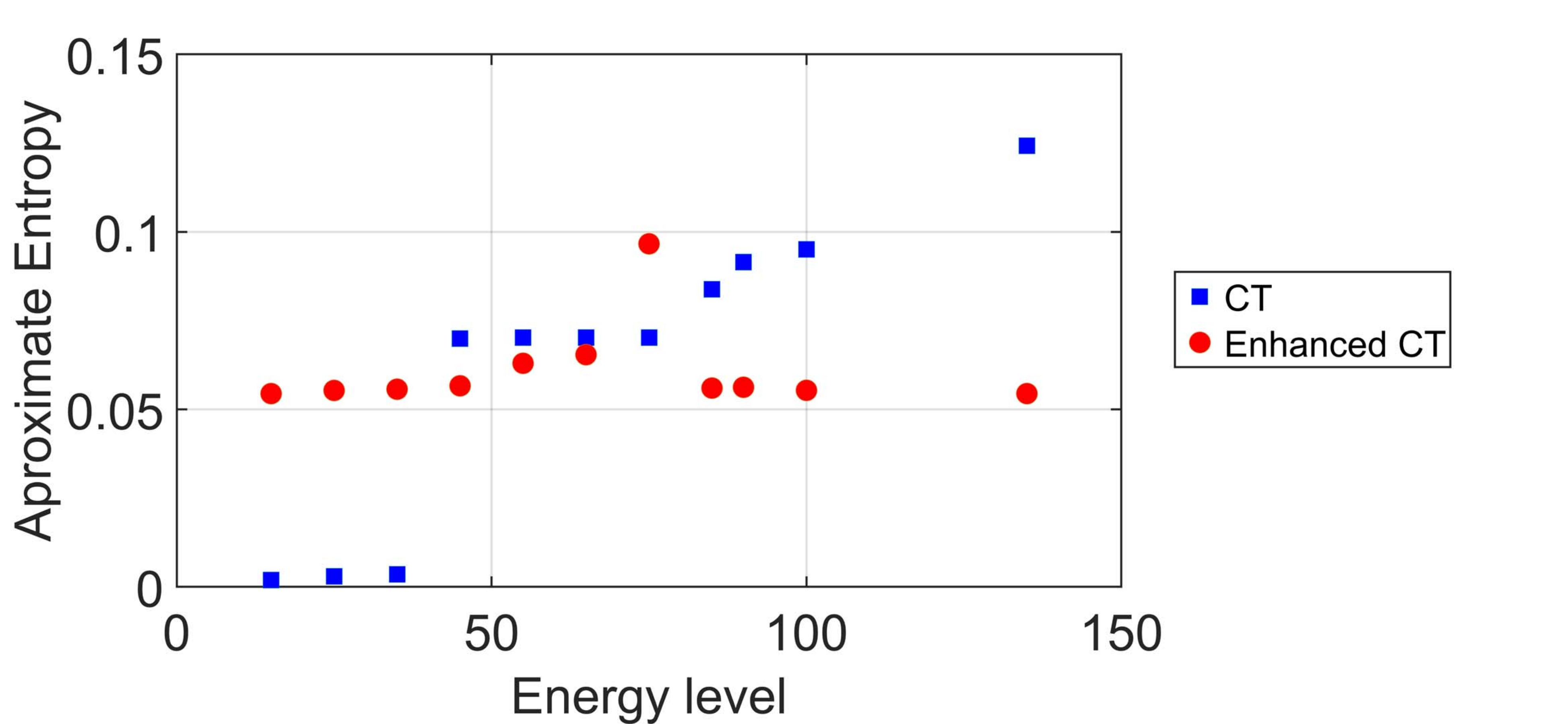}
	\caption{Approximate entropy of CT and enhanced CT images in different energy levels.}\label{fig:12}
\end{figure}

\begin{figure}[H]
	\centering
	\includegraphics[width=1\textwidth]{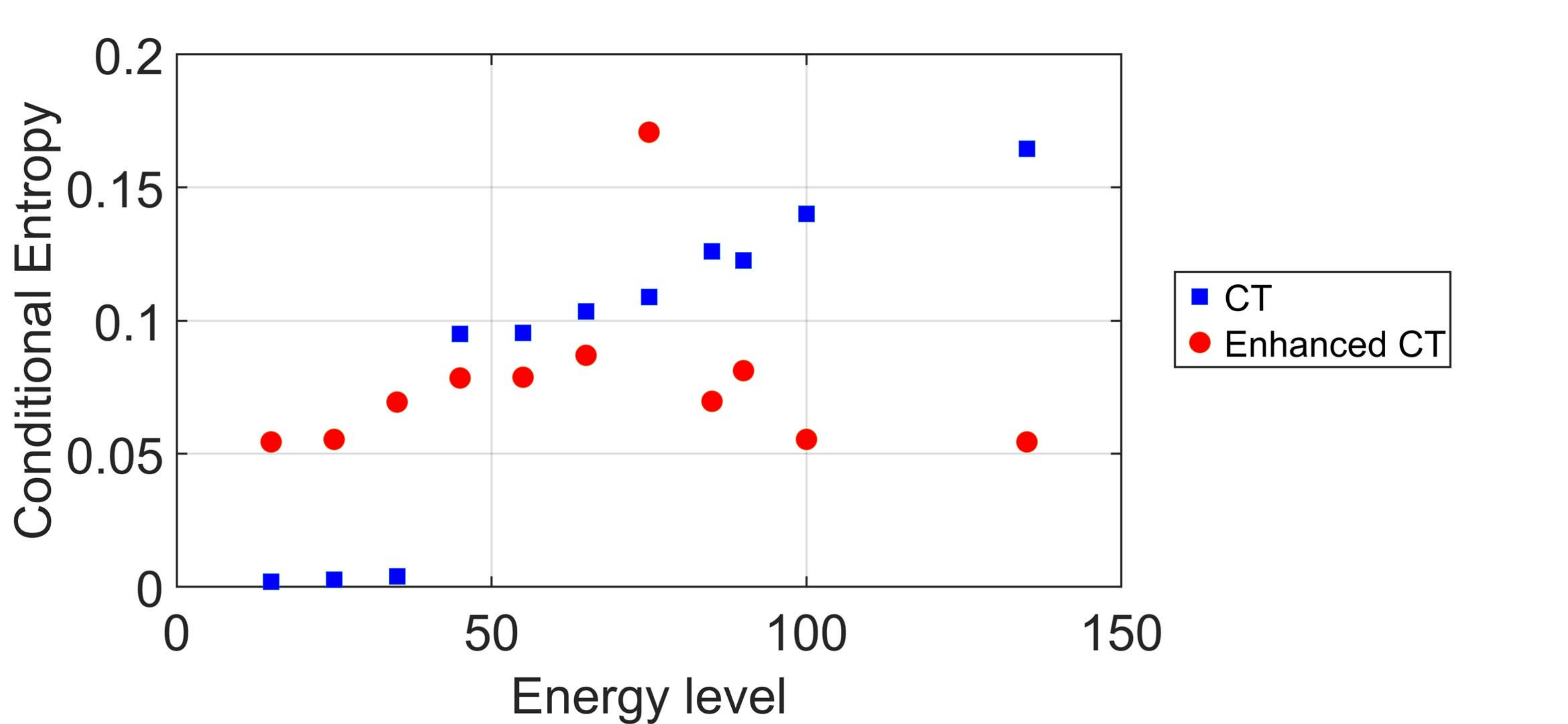}
	\caption{Conditional entropy of CT and enhanced CT images in different energy levels.}\label{fig:13}
\end{figure}

Approximate entropy (ApEn) \cite{pincus1991regularity} quantifies the amount of regularity and the unpredictability of fluctuations over time-series data. It modifies an exact regularity statistic, i.e., Kolmogorov-Sinai entropy, to handle the system noise when the amounts of data are not vast enough and the study deals with the experimental data. Results of calculating ApEn for both CT and enhanced CT images are illustrated in Fig.~ \ref{fig:12}. We stated that the quantizing energy levels is done by taking Kolmogorov-Smirnov test to find the best distribution fits the conditional entropy. Therefore, CT images conditioned on the known energy levels and quantifying the amount of information needed to describe the outcome of CT images can be better done by measuring Conditional Entropy \cite{cover2012elements} (see Fig.~ \ref{fig:13}). Given discrete random variables $X$ with image $\mathcal X$ and $Y$ with image $\mathcal Y$, the conditional entropy is defined by Eq. \ref{eq:7}.

\begin{equation}\label{eq:7}
H(Y|X)\ \equiv\sum _{{x\in {\mathcal  X},y\in {\mathcal  Y}}}p(x,y)\log {\frac  {p(x)}{p(x,y)}}
\end{equation}

\begin{figure}[H]
	\centering
	\includegraphics[width=1\textwidth]{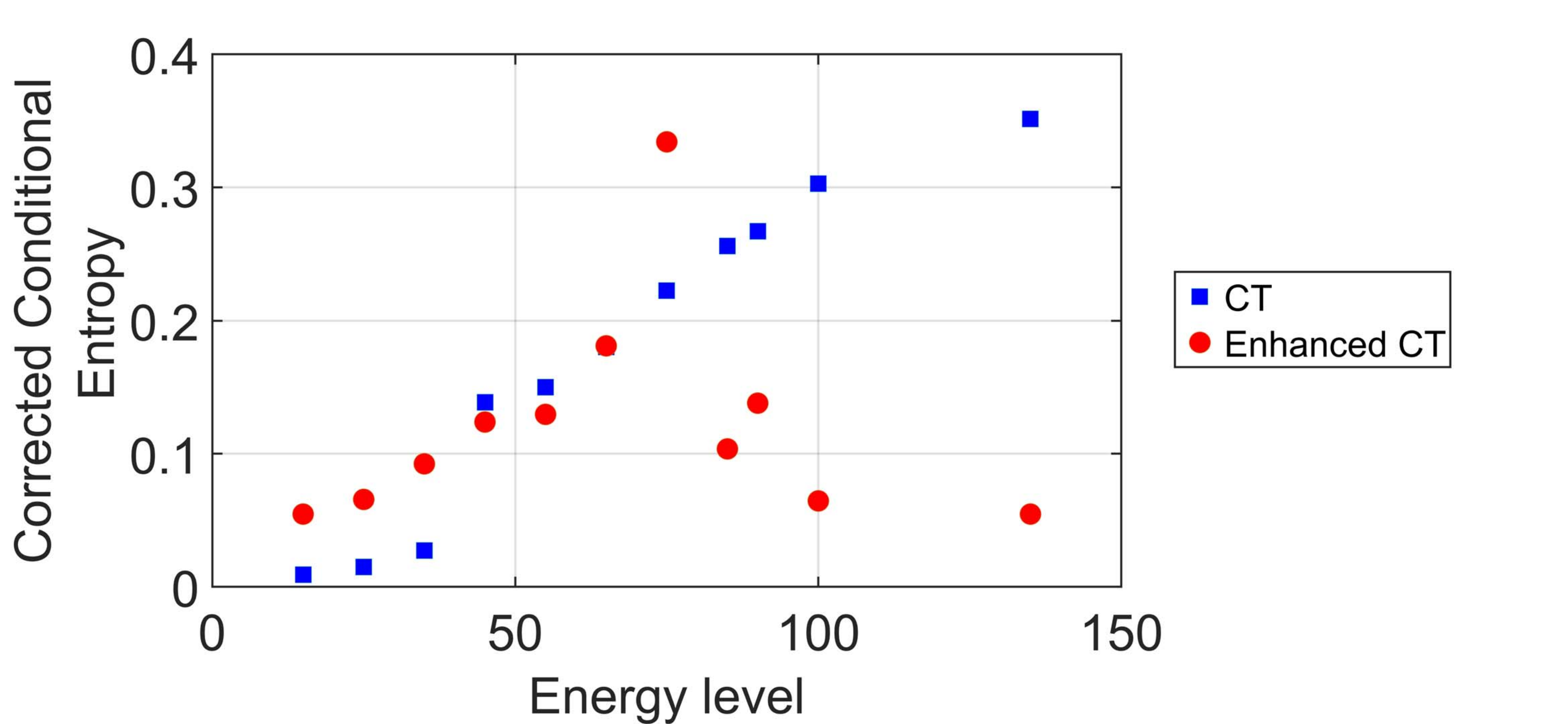}
	\caption{Corrected Conditional entropy of CT and enhanced CT images in different energy levels.}\label{fig:14}
\end{figure}

Limited number of samples leads to the growing percentage of single points in $L$-dimensional phase space when $L$ increases which subsequently increase the probability of the a-priori selection of the embedding dimension. To handle the mentioned problems, one could use Corrected Conditional Entropy (CCE) in which the information content can be measured based on the search for the minimum of the defined function in Eq. \ref{eq:8}. This value is taken as an index in the information domain quantifying the regularity of the process and experienced an increase when no robust statistic can be performed as a result of a limited amount of available samples.

\begin{equation}\label{eq:8}
\begin{split}
CCE(L) = \hat{E}(L/l-1) + E_{c}(L)\\
E_{c}(L)= \mathrm{perc}(L) . \hat{E}(1)
\end{split}
\end{equation}

where $\hat{E}(L/l-1)$ represent the estimate of Shannon entropy (SE) in a $L/L-1$-dimensional phase space, $\mathrm{perc}(L)$ is the percentage of single points in the $L$-dimensional phase space, and $\hat{E}(1)$ the estimated value of SE for $L = 1$. Figure~ \ref{fig:14} shows CCE with its the energy level components calculated over CT and enhanced CT images. The entropy change rate of eCT images is lower than CT images while the energy level increases. It is remarkable that the CCE values increase in all of CT images while it experiences a lower change in eCT images.

Although the variations rhythm in both Fig.~ \ref{fig:13} and \ref{fig:14} seems to be homogeneous, one can see the conditional entropy values of the CT images in energy levels of 15, 25 and 35 are near zero. Therefore, it is likely to infer that reconstructed CT images in these energy levels are certainty determinable, and far from stochastic conditions. This high degree of certainty contradicts the nature of medical imaging, where three different tissues were considered within the phantom. Corrected conditional entropy, however, resolves this issue by considering the mentioned assumptions and covers the associated problem with the low number of data.

\begin{figure}[H]
	\centering
	\includegraphics[width=1\textwidth]{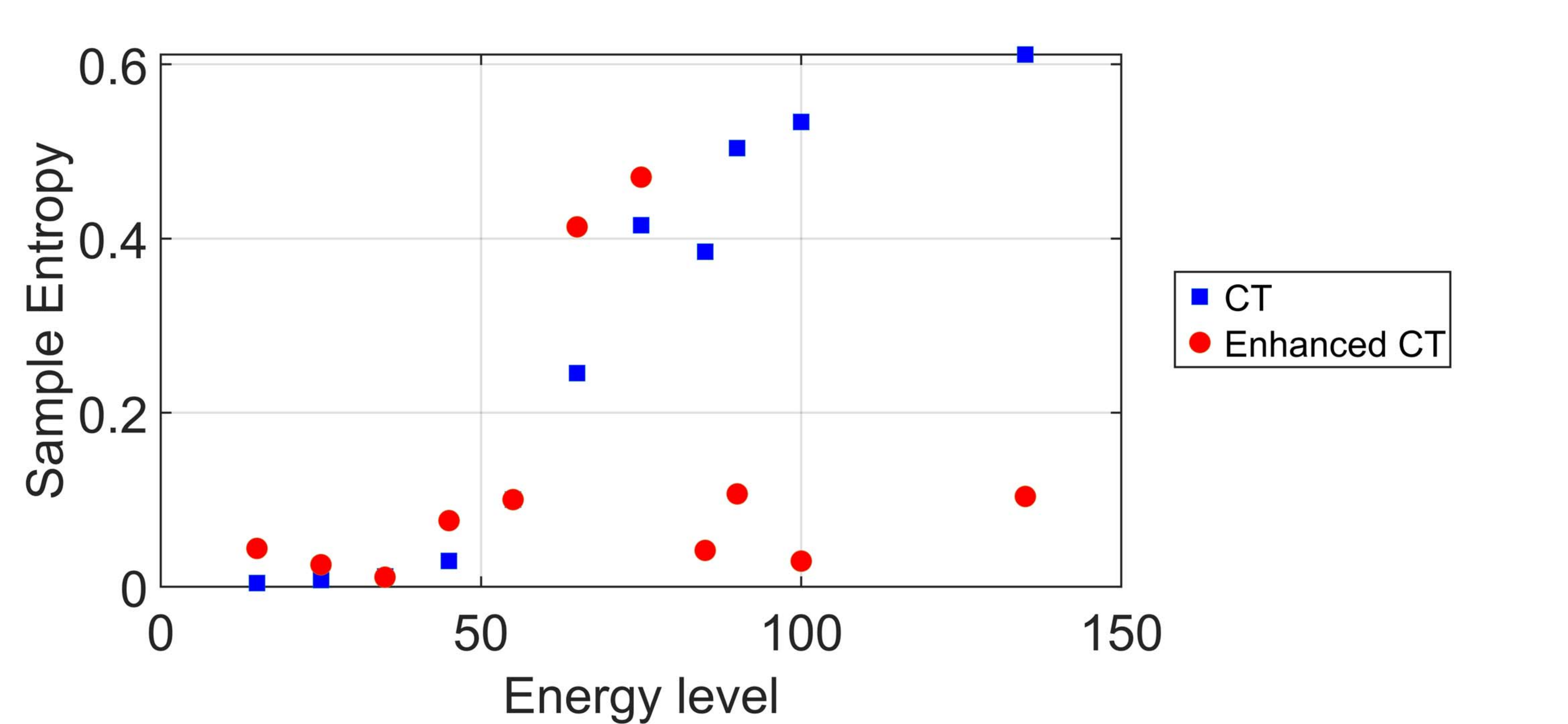}
	\caption{Sample entropy of CT and enhanced CT images at different energy levels.}\label{fig:15}
\end{figure}

Sample entropy (SampEn) \cite{richman2000physiological}, a measure of complexity, is a modification of approximate entropy with two advantages over ApEn including independence of data length and a relatively trouble-free implementation. As self-matching is not included in SampEn, actual interpretation about the irregularity of signals is more possible. For a given embedding dimension $m$, tolerance $r$ and number of data points $N$, SampEn is calculated by Eq. \ref{eq:9}.

\begin{equation}
\label{eq:9}
\mathrm{SampEn} = - \log \frac{A}{B},
\end{equation}

where $A$ is a number of template vector, of length $m+1$, pairs such as $d[X_{{m+1}}(i),X_{{m+1}}(j)]<r$  and $B$ is a  number of template vector,of length $m$, pairs such as $d[X_{m}(i),X_{m}(j)]<r$. Figure~ \ref{fig:15} shows the results of calculating SampEN for both CT and enhanced CT images.

\begin{figure}[H]
	\centering
	\includegraphics[width=1\textwidth]{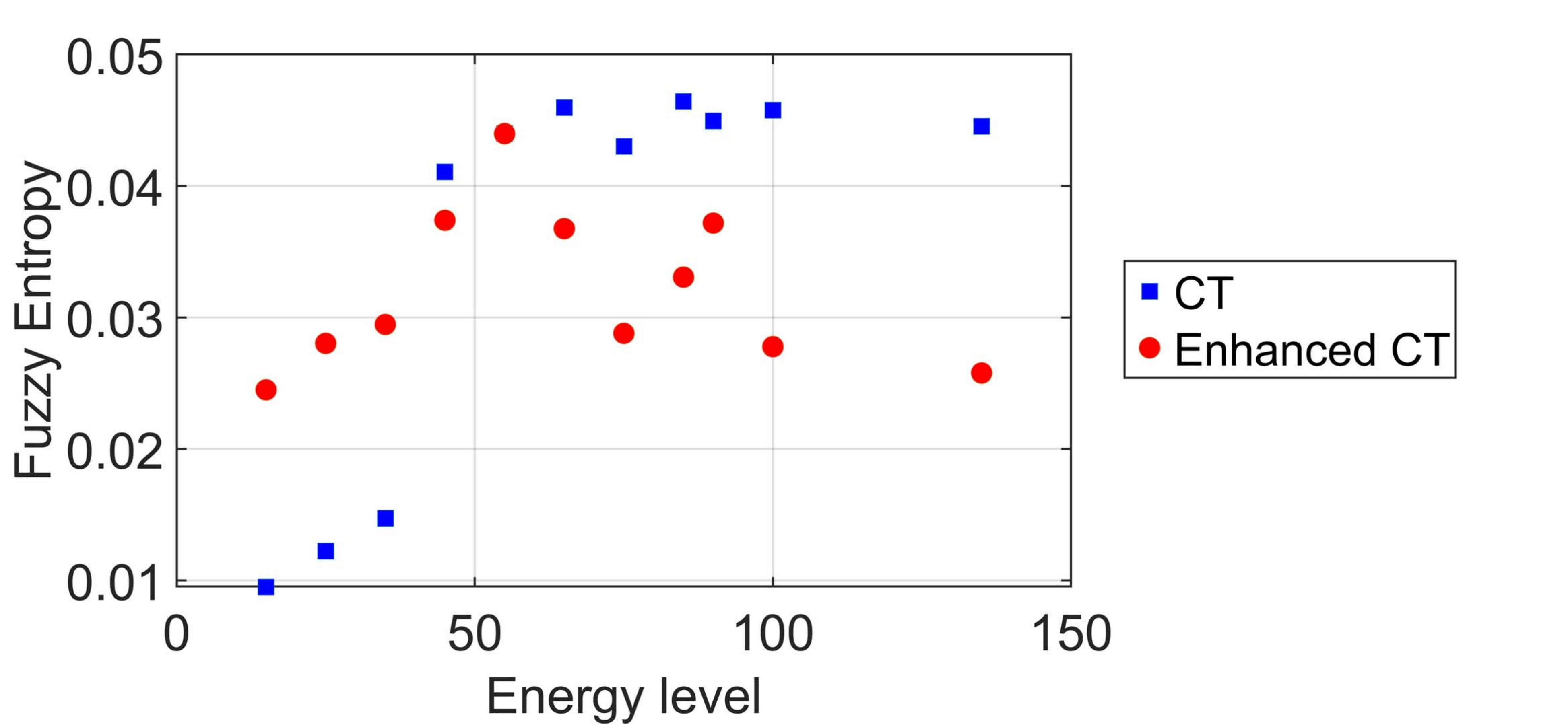}
	\caption{Fuzzy entropy of CT and enhanced CT images in different energy levels.}\label{fig:16}
\end{figure}

Fuzzy entropy (FuzzyEn) can be used in analyzing nonlinear time series using modified sample entropy. FuzzyEn well estimates the short data where its validity is not restricted by the parameter value. This measure evaluates global deviations from the type of ordinary sets. Furthermore, it is resistant to noise and jamming phenomena. FuzzyEn can be defined for a given time series by using Eq. \ref{eq:10}.

\begin{equation}
\begin{split}
FuzzyEn (m, n, r, N) = \ln \phi^{m} (n, r) - \ln \phi^{m+1} (n, r), \\
\phi (n,r) = \frac{1}{N-m} \sum_{i=1}^{N-m} [\frac{1}{N-m-1} \sum_{j=1, j \neq i}^{N-m} D_{ij}^{m}]
\end{split}
\label{eq:10}	
\end{equation}

where $m$ and $r$ are the dimensions of phase space and similarity tolerance, respectively, $n$ is the gradient of the exponential function, $N$ is the number of data, and $D$ is the similarity degree. Figure~\ref{fig:16} shows the results of calculating FuzzyEn for both CT and enhanced CT images.

The last entropy measure is Permutation entropy (PE) which can consider neighboring values of data in its measuring. This complexity measure is most appropriate for chaotic time series, in particular in the presence of dynamical and observational noise. As a small noise does not essentially change the complexity of a chaotic signal, PE behaves similarly to Lyapunov exponents where it is known as a complexity parameter. Considering a time series $\{x_{t}\}_{t=1, \cdots, T}$, one could study all $n!$ permutations $\pi$ of order $n$ and determine the relative frequency by Eq. \ref{eq:11}.

\begin{equation}
p(\pi) = \frac{\mathbf{card}(\{t | t \leq T-n, (x_{t+1}, \cdots, x_{t+n}) \quad \mathbf{has type} \quad \pi\})}{T-n+1}
\label{eq:11}	
\end{equation}

\begin{figure}[H]
	\centering
	\includegraphics[width=1\textwidth]{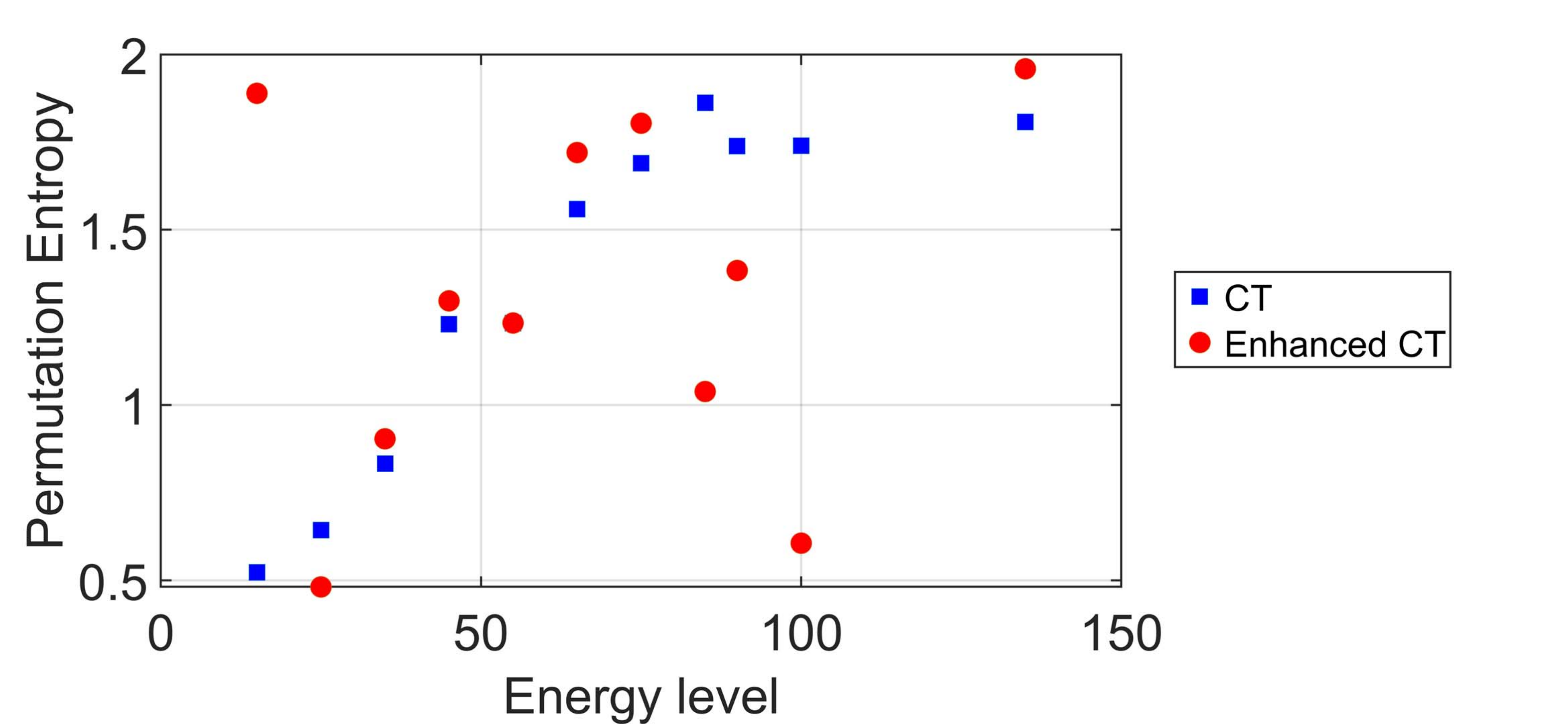}
	\caption{Permutation entropy of CT and enhanced CT images in different energy levels.}
	\label{fig:17}
\end{figure}

The permutation entropy of order $n \geq 2$ is defined as Eq. \ref{eq:12}.

\begin{equation}
H(n) = -\sum p(\pi)\log p(\pi)
\label{eq:12}	
\end{equation}

where the sum runs over all $n!$ permutations $p$ of order $n$ and $n$ is the dimension of data. Results of calculating PE for Ct and enhanced CT images are shown in Fig.~ \ref{fig:17}. 

In all the measured entropies, we see that irregularity of CT images raises with increasing energy levels, whereas enhanced CT images have lower, yet more tolerant, increasing rates and in some cases, they have a dual behavior. Therefore, it is likely that analyzing enhanced CT images can produce more reliable results. This claim is what we will investigate it by performing a morphological richness analysis \cite{taghipour2016complexity} as well as Fuzzy C-means (FCM) \cite{gan2007data} based segmentation.

Morphological richness (MR) is calculated as the number of different configurations of $3 \times 3$ blocks divided by the number of all possible configurations $(2^{9})$. Although the results must give us a deep sense about the restructuring of reconstructed images by different energies, the chaotic nature of each data leaves us far from the desired inference. Hence, the power spectrum of the calculated morphological richness is illustrated to make the complexity analysis sensible. To this end, the Fourier transform is applied to MR to swap the dimension of time with the dimension of frequency. A very strong and slow component in the frequency domain implies that there is a high correlation between the large-scale pieces of the signal in time (macro-structures), while a very strong and fast oscillation implies correlation in the micro-structures. Therefore, if our signal $f(t)$ represents values in every single moment of time, its Fourier transform $F(\omega)$ represents the strength of every oscillation in a holistic way in that chunk of time. These two signals are related to each other by Eq. \ref{eq:13}:

\begin{equation}
F(\omega) = \int_{-\infty}^{ \infty} f(t)e^{-j \omega t} dt,
\label{eq:13}	
\end{equation}

Limitation on not always being able to observe a signal from $-\infty$ to $\infty$ causes defining $F_{T}(\omega)$ in period $T$. In this way, the power spectrum is calculated by Eq. \ref{eq:14}.

\begin{equation}
S_{f}(\omega) = \lim_{T \rightarrow \infty} \frac{1}{T}|F_{T}(\omega)|^{2}.
\label{eq:14}
\end{equation}

The power spectrum itself is the Fourier transform of the autocorrelation function. The autocorrelation function represents the relationship of long- and short-term correlation within the signal itself (refer to Eq. \ref{eq:15}).

\begin{equation}
<f(t), f(t+\tau)> = \frac{1}{2\pi}\int_{0}^{ \infty}S_{f}(\omega)f(t)e^{-j \omega t} d\omega
\label{eq:15}
\end{equation}

The results of our analysis are illustrated in Fig.~ \ref{fig:18}. Amplitude and \enquote{dominating frequencies} differentiations are evident in enhanced CT images which imply that analyzing enhanced CT images would bring more information.

\begin{figure}[H]
	\centering
	\begin{subfigure}[b]{0.9\textwidth}
		\includegraphics[width=\textwidth]{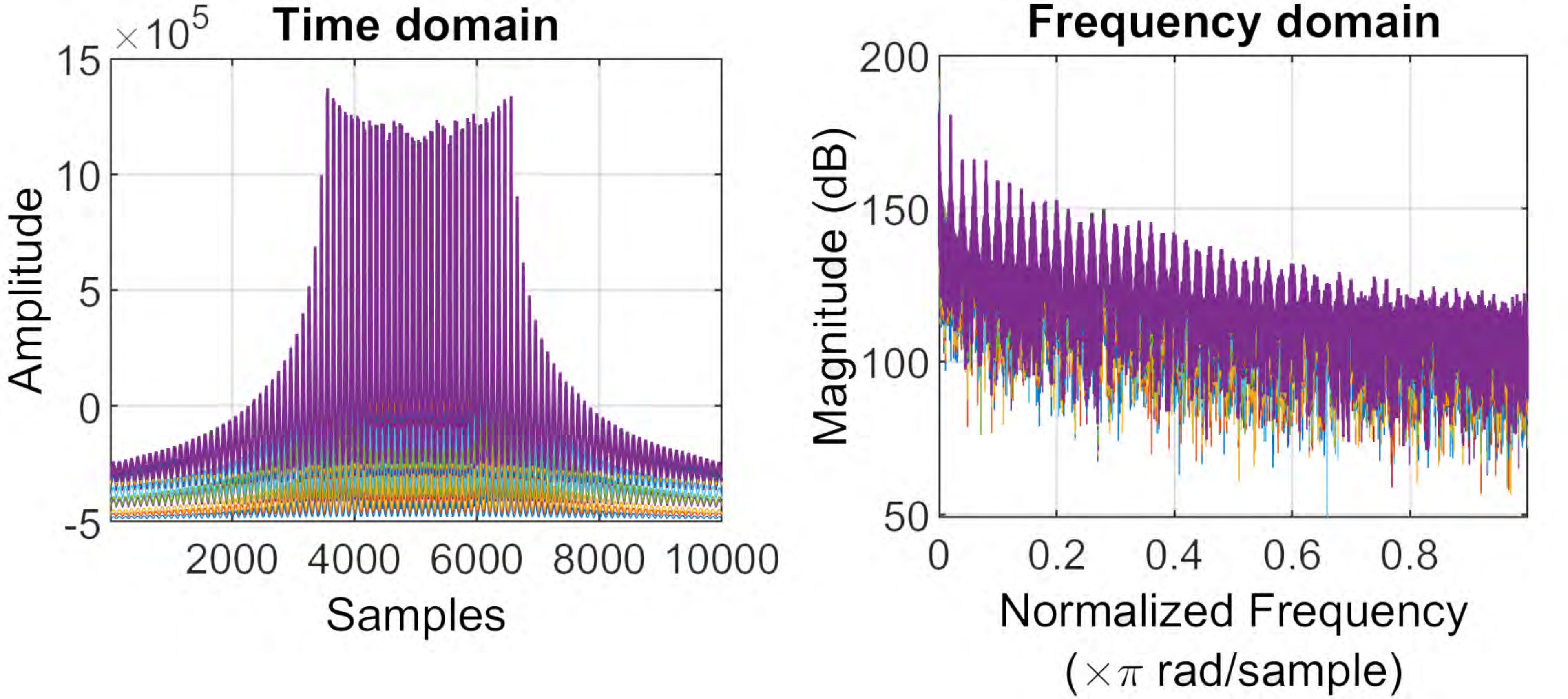}
		\caption{}
	\end{subfigure}
\end{figure}
\begin{figure}[H]\ContinuedFloat
	\begin{subfigure}[b]{0.9\textwidth}
		\includegraphics[width=\textwidth]{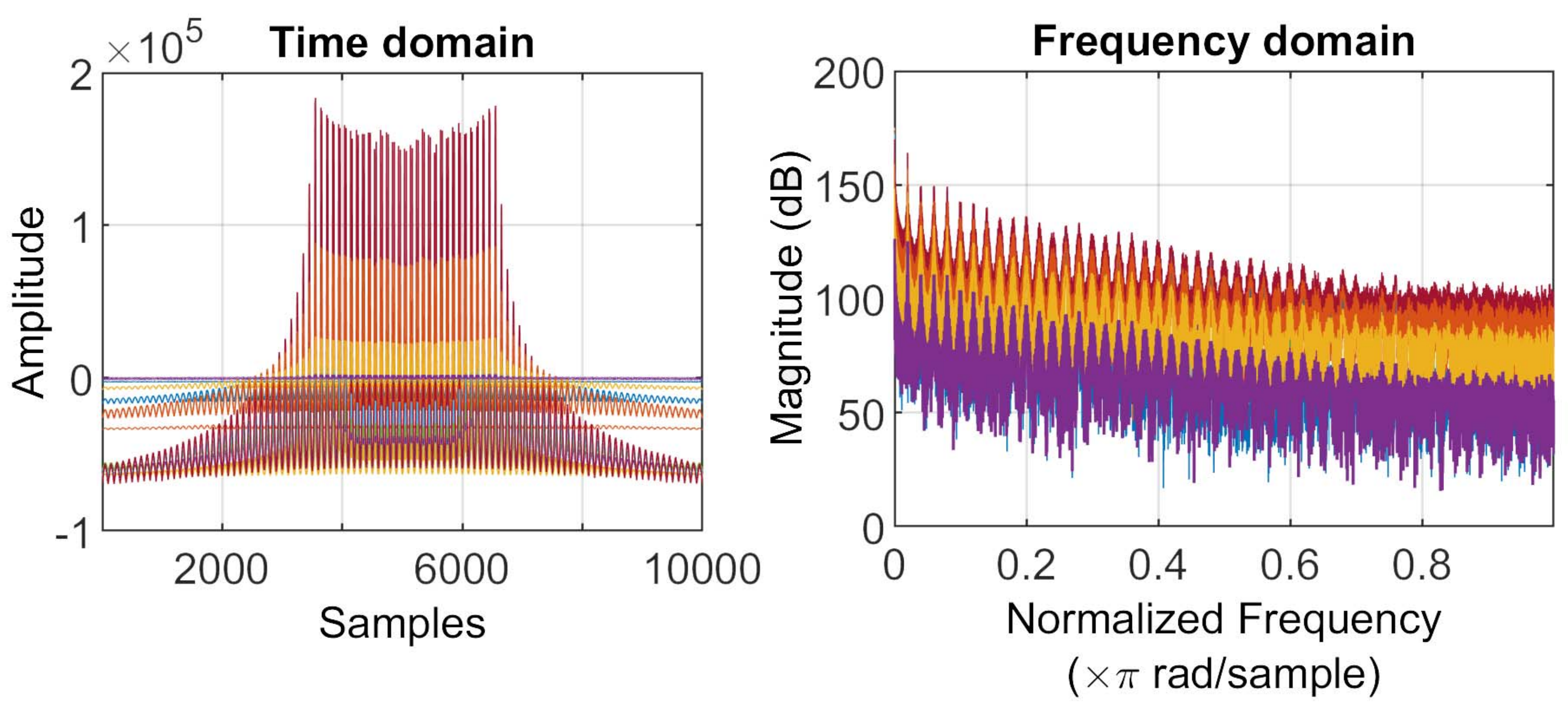}
		\caption{}
	\end{subfigure}
	\caption{Power spectrum of the entropy of the calculated morphological richness.(a) CT images, (b) enhanced CT images.}\label{fig:18}
\end{figure}

Image segmentation plays an important role in medical image processing \cite{dehshibi2017hybrid}. Fuzzy c-means (FCM) is one of the popular clustering algorithms \cite{gan2007data} for medical image segmentation. But FCM is highly vulnerable to noise due to not considering the spatial information in image segmentation. Therefore, we investigate how much FCM is resistant against artefacts when applies to the enhanced CT images. FCM minimizes an object function by partitioning a finite collection of $n$ elements $X=\{{\mathbf  {x}}_{1},...,{\mathbf  {x}}_{n}\}$ into a collection of c fuzzy clusters with respect to some given criterion. FCM returns a list of $c$ cluster centers $C=\{{\mathbf  {c}}_{1},...,{\mathbf  {c}}_{c}\}$ and a partition matrix $W=w_{{i,j}}\in [0,1],\;i=1,...,n,\;j=1,...,c$, where each element, $w_{ij}$ , tells the degree to which element, $\mathbf {x} _{i}$, belongs to cluster ${\mathbf  {c}}_{j}$. The objective function can be defined by Eq. \ref{eq:16}

\begin{equation}
\label{eq:16}
\begin{split}
{\underset  {C}{\operatorname {arg\,min}}}\sum _{{i=1}}^{{n}}\sum_{{j=1}}^{{c}}w_{{ij}}^{m}\left\|{\mathbf  {x}}_{i}-{\mathbf  {c}}_{j}\right\|^{2}, \\
w_{{ij}}={\frac  {1}{\sum _{{k=1}}^{{c}}\left({\frac  {\left\|{\mathbf  {x}}_{i}-{\mathbf  {c}}_{j}\right\|}{\left\|{\mathbf  {x}}_{i}-{\mathbf  {c}}_{k}\right\|}}\right)^{{{\frac  {2}{m-1}}}}}}.
\end{split}
\end{equation}

\begin{figure}[H]
	\centering
	\begin{subfigure}[b]{0.48\textwidth}
		\includegraphics[width=\textwidth]{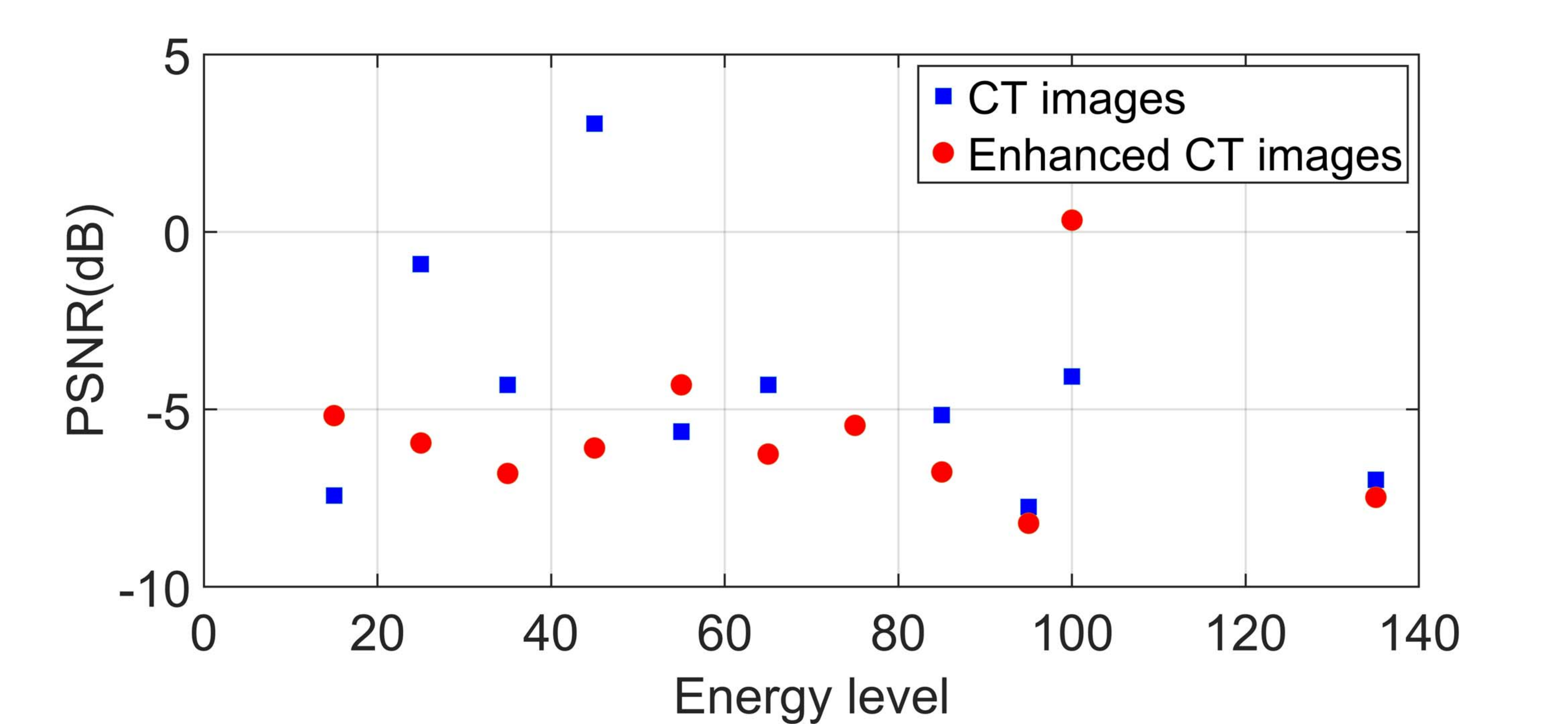}
		\caption{}
	\end{subfigure}
	\hfill
	\begin{subfigure}[b]{0.48\textwidth}
		\includegraphics[width=\textwidth]{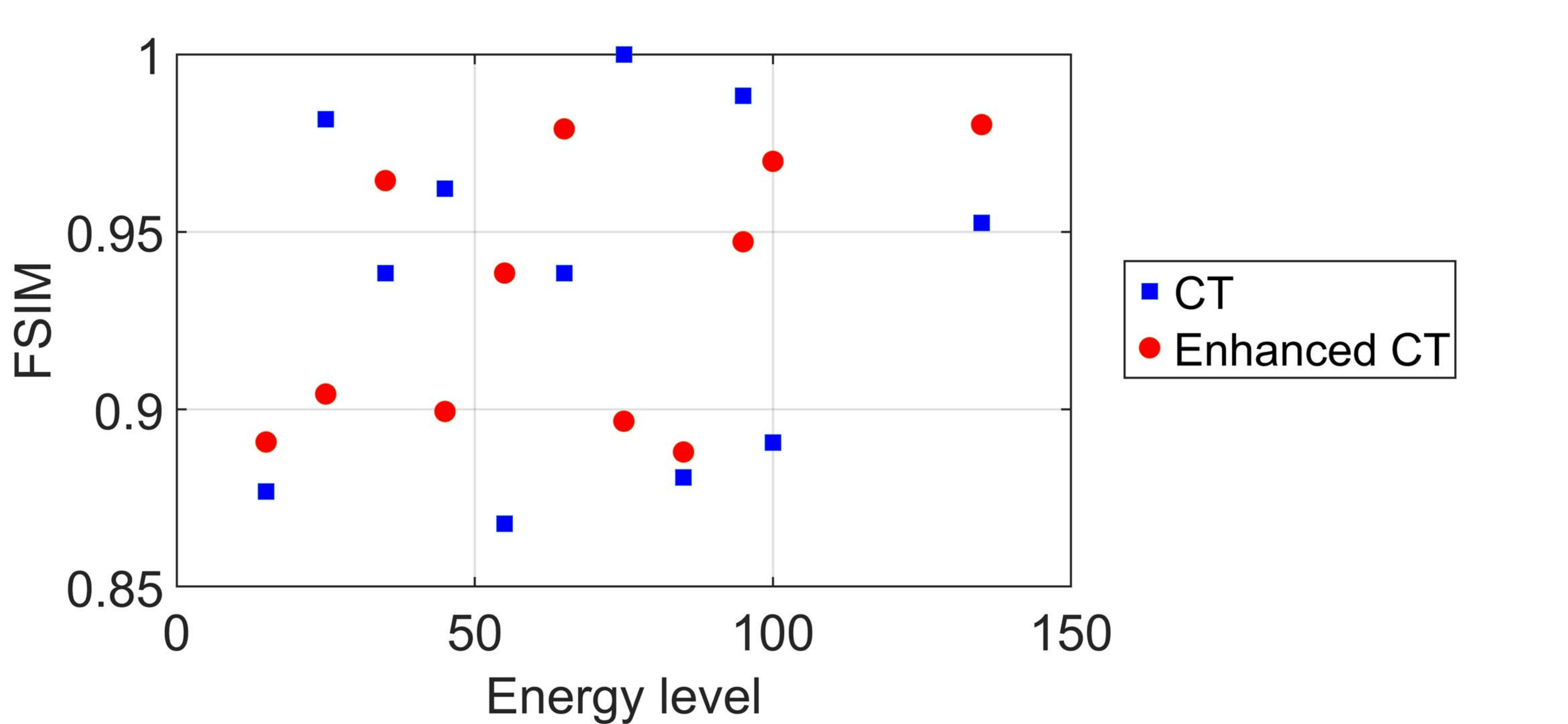}
		\caption{}
	\end{subfigure}
	\hfill
	\begin{subfigure}[b]{0.48\textwidth}
		\includegraphics[width=\textwidth]{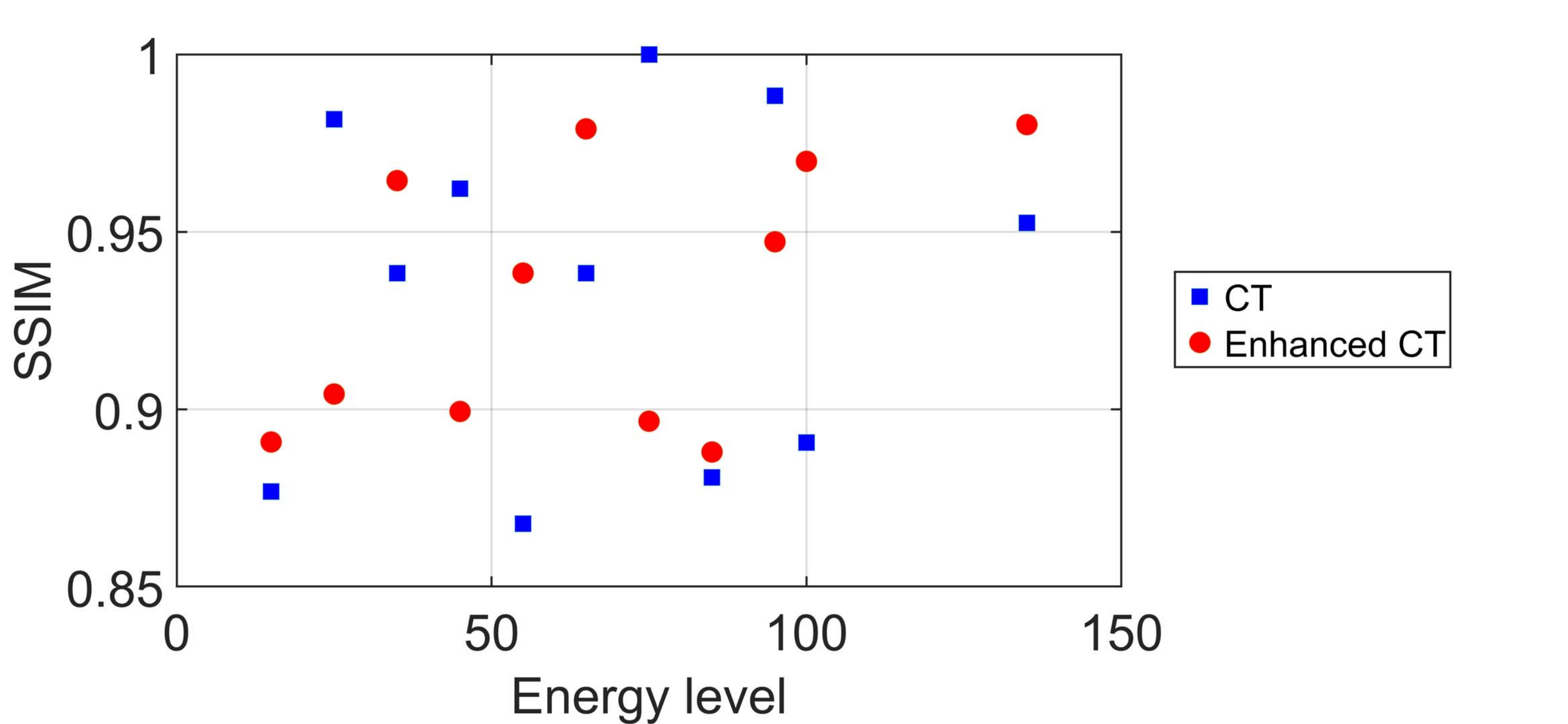}
		\caption{}
	\end{subfigure}
	\hfill
	\begin{subfigure}[b]{0.48\textwidth}
		\includegraphics[width=\textwidth]{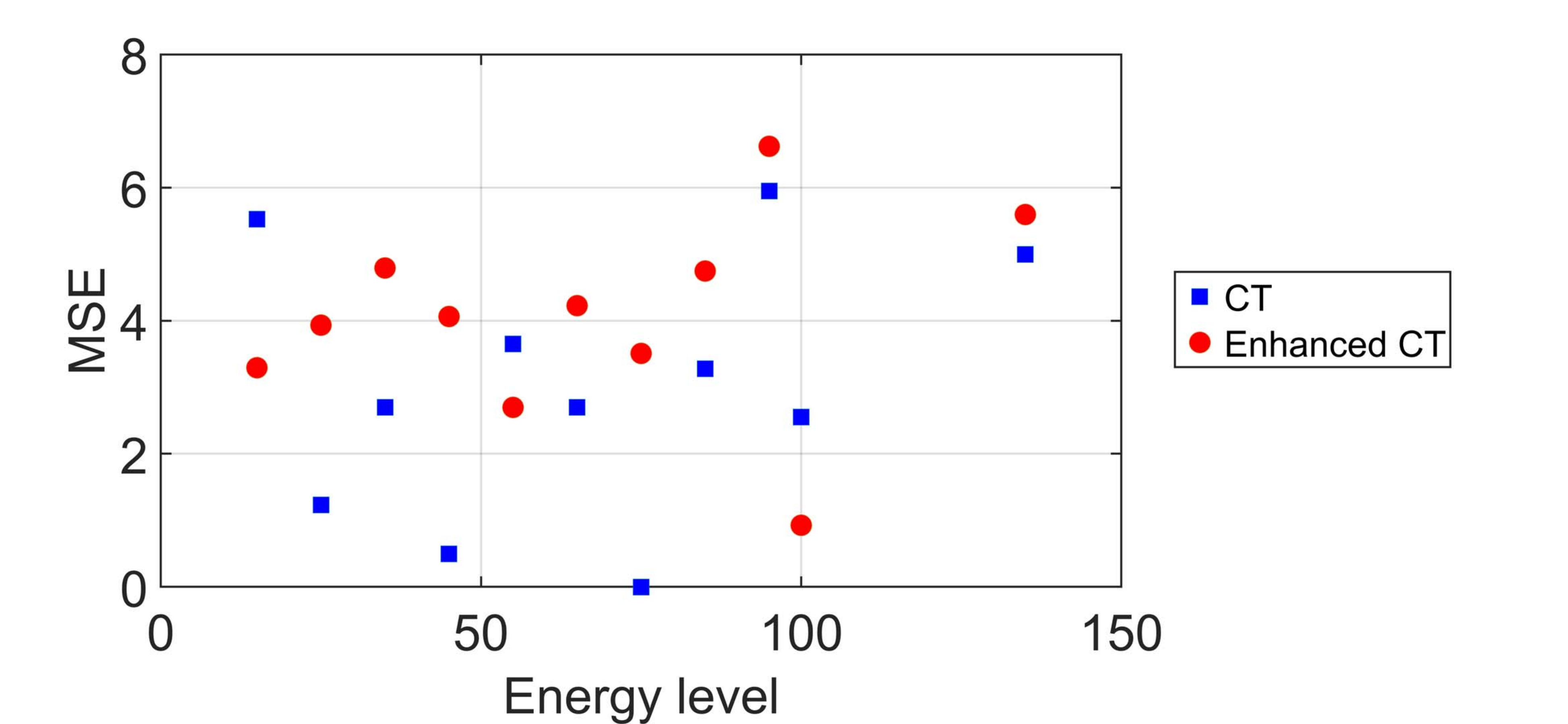}
		\caption{}
	\end{subfigure}
	\caption{Evaluation criteria to analysis the FCM applied to both CT and enhanced CT images.(a) PSNR, (b) FSIM, (c) SSIM, and (d) MSE.}\label{fig:19}
\end{figure}

Peak-value signal-to-noise ratio (PSNR), feature-similarity (FSIM) index, Structural Similarity (SSIM) index, and Mean Square Error (MSE) are chosen as the evaluation criteria (refer to Eq. \ref{eq:17}).

\begin{gather}
\label{eq:17}
{\mathrm{PSNR}(I_{input}, I_{reference})} = 10\cdot \log_{{10}}\left({\frac{{\mathit{MAX}}_{I_{input}}^{2}}{{\mathit  {MSE}}}}\right), \\
{\mathrm{FSIM}(I_{input}, I_{reference})} = \frac{\sum_{x \in \Omega}S_{L}(x) \cdot PC_{m}(x)}{\sum_{x \in \Omega}PC_{m}(x)}, \\
{\mathrm{SSIM}}(I_{input}, I_{reference})={\frac {(2\mu_{I_{input}}\mu_{I_{reference}}+c_{1})(2\sigma_{{I_{input}\cdot I_{reference}}} + c_{2})}{(\mu_{I_{input}}^{2} + \mu_{I_{reference}}^{2} + c_{1})(\sigma_{I_{input}}^{2} + \sigma_{I_{reference}}^{2} + c_{2})}}, \\
{\mathrm{MSE}(I_{input}, I_{reference})}={\frac{1}{m\,n}}\sum _{{i=0}}^{{m-1}}\sum _{{j=0}}^{{n-1}}[I_{input}(i,j)-I_{reference}(i,j)]^{2}.
\end{gather}

where $I_{input}$ is the image with the size of $m \times n$, $PC_{m}$ is the weighting factor for $S_{L}(x)$ which is the overall similarity between $I_{input}$ and a reference image $I_{reference}$, $\mu$ is the average of the image, and $\sigma$ is the variance of image. Plots of calculated measures are illustrated in Fig.~\ref{fig:19}. 

Based on the evidence in Figs. \ref{fig:6}-\ref{fig:19}, our findings confirm the following:

\begin{enumerate}
	\item Energy levels between 50-90~keV show abnormal changes in entropy measures for both CT and enhanced CT images. This is because the resultant tissues produced in the phantom have the least tissue differentiation than the water has.
	\item PSNR plot of enhanced CT images shows that the variation in tolerances of the enhanced CT images is less than that of CT images. Therefore, one can conduct different analyzes at various energy levels with less concern.
	\item In the majority of energy levels, it is obvious that PSNR, FSIM, SSIM, and MSE report for better values in comparison of enhanced CT images with the conventional CT image. Therefore, it is reasonable to conclude that with a lower degree of irradiation and by applying the proposed post-processing method, one can reach to a better discrimination in analyzing images whereas this less irradiation causes less tissue damage.
	\item It is an accepted fact that reconstructing CT images from the constructed Sinogram has to be done in the energy level of 70 keV. Results of our experiments prove that it is possible to confidently work on CT images in different energy levels by applying either the proposed post-processing method or physical modification. In this way, an expert can reach to a better CT image where the objective tissue is more discriminative in comparison with the surrounded tissues.
\end{enumerate}

\section{Conclusion}
We presented a method of modifying reconstructed CT image in GATE/GEANT4 environment using the applying weights of photon flux. This post-processing method will contribute toward analyzing CT images by easing the computational inference about different tissues irradiated in different energy levels. Our evaluations of generative complexity might open several new venues in medical imaging. These are related to the complexity hierarchies of the CT images and the relation of the complexity hierarchies to the enhancing of these images. The morphological richness along with entropy can derive a one-tone mapping among the evolution of tissues irradiated in different energy levels with respect to the water attenuation map. This analysis can then be used as a tool in a predictive technique for forecasting future developments in the medical imaging task.

The proposed method consists of several main steps including (1) back-projecting acquired data to form pixel-based attenuation matrix (PAM); (2) finding the statistical average of each interval to use as the effective energies; (3) calculating HU scale of each interval (4) computing the associated photon fluxes based on X-ray spectrum; (5) modifying HU scales by weighting them with the computed fluxes. Visual and complexity analysis convince us to touch on the topic of non-constructability. In cellular automaton theory \cite{wolfram1984cellular}, \cite{myhill1963converse} a configuration is called nonconstructable or Garden-of-Eden if it could not be reached from any other configuration by applying local rules of cell state transitions. When adopting the concept in the reconstructing of CT images by modification with photon flux, we can talk about a degree of nonconstructability; the bigger the relative complexity of an image, the higher the degree of non-constructability. We can hypothesize that the higher the degree of non-constructibility of a CT image, the most discriminating the target tissue emerged in the constructed phantom.

\bibliographystyle{elsarticle-num}
\bibliography{bibliography}

\end{document}